\documentclass[
    prb,
    reprint,
    superscriptaddress,
    floatfix,
    longbibliography
]{revtex4-2}

\usepackage{anyfontsize}
\usepackage{lmodern}

\usepackage{amssymb}
\usepackage{amsmath}
\usepackage{amsfonts}
\usepackage{bm}
\usepackage{bbm}
\usepackage{braket}

\newcommand{\pqty}[1]{\left( #1 \right)} 
\newcommand{\bqty}[1]{\left[ #1 \right]} 
\newcommand{\ev}[1]{\langle #1 \rangle} 
\newcommand{\abs}[1]{\left| #1 \right|} 

\usepackage{graphicx}

\usepackage{xcolor}
\colorlet{Red}{red!70!black}
\colorlet{Blue}{blue!70!black}
\colorlet{Gray}{gray!70!black}
\colorlet{Green}{green!70!black}

\usepackage[colorlinks, allcolors=blue]{hyperref}
\usepackage[caption=false, position=top, margin=-5pt]{subfig}

\renewcommand{\paragraph}[1]{\textbf{#1}.~~}

\renewcommand{\L}{\mathbb{L}} 
\newcommand{\Z}{\mathbb{Z}} 
\newcommand{\One}{\mathbbm{1}} 
\newcommand{\Hphys}{\mathcal{H}_{\text{phys}}} 
\newcommand{\toplink}{\ell^\uparrow}
\newcommand{\botlink}{\ell^\downarrow}
\newcommand{\runglink}{\ell^0}
\newcommand{\Vup}{V^{\uparrow}}
\newcommand{\Vdown}{V^{\downarrow}}
\newcommand{\Vh}{V^0}
\newcommand{\coeffup}{c^{\uparrow}}
\newcommand{\coeffdown}{c^{\downarrow}}

\newcommand{\Unibo}{Dipartimento di Fisica e Astronomia, Universit\`a di Bologna, 40127 Bologna, Italy}
\newcommand{\Unifi}{Dipartimento di Fisica e Astronomia, Universit\`a di Firenze, 50019 Firenze, Italy}
\newcommand{\InfnBo}{INFN, Sezione di Bologna, I-40127 Bologna, Italy}
\newcommand{\InfnFi}{INFN, Sezione di Firenze, I-50019 Firenze, Italy}

\usepackage{pgfplots}
\pgfplotsset{compat=1.16} 

\begin{document}

\title{Discrete Abelian lattice gauge theories on a ladder \texorpdfstring{\\}{}and their dualities with quantum clock models}

\author{S.~Pradhan}
\altaffiliation{These authors have contributed equally.}
\affiliation{\Unibo}
\affiliation{\InfnBo}

\author{A.~Maroncelli}
\altaffiliation{These authors have contributed equally.}
\affiliation{\Unifi}
\affiliation{\InfnFi}

\author{E.~Ercolessi}
\affiliation{\Unibo}
\affiliation{\InfnBo}

\begin{abstract}
    We study a duality transformation from the gauge-invariant subspace of a $\Z_N$ lattice gauge theory on a two-leg ladder geometry to an $N$-clock model on a single chain.
    The main feature of this mapping is the emergence of a longitudinal field in the clock model, whose value depends on the superselection sector of the gauge model, implying that the different sectors of the gauge theory can show quite different phase diagrams.
    In order to investigate this and see if confined phases might emerge, we perform a numerical analysis for $N = 2, 3, 4$, using both exact diagonalization and DMRG.
\end{abstract}

\maketitle

\section{Introduction}
\label{sec:introduction}

Gauge theories constitute the baseline in our microscopical description of physical fundamental laws and are a cornerstone of contemporary scientific research.
Calculations beyond perturbative regimes, as needed to understand for example the quark confinement mechanism in Quantum Chromodynamics, represent a notorious challenge both analytically and numerically.
Standard classical computational methods adopt the Wilson's framework of lattice gauge theories (LGTs) \cite{wilson1974confinement, kogut1975hamiltonian, kogut1979introduction}, in which the continuous space–time is replaced by a discrete set of points and the calculations are performed in the Euclidean path-integral approach.
More recently, inspired by Feynman's idea of quantum simulations \cite{feynman2018simulating, feynman1985quantum}, many authors have adopted a Hamiltonian approach in which only spatial coordinates are discretized, and which might be implemented via a quantum platform once the group degrees of freedom are also discretized, by considering a finite group or by suitable approximations (see \cite{banuls2020simulating, zohar2015quantum, dalmonte2016latticegauge, zohar2015latticegauge, zohar2017digital} and references therein).
Still, in all approaches, enforcing the gauge constraints to restrict the (analytical, numerical or experimental) evaluation of observables to the gauge-invariant Hilbert subspace is a challenging task, which is dealt with different strategies.

In this paper we consider Abelian LGTs, which are known to exhibit confined/deconfined phases \cite{fradkin1979phase, horn1979zngauge, tagliacozzo2011entanglement, tagliacozzo2013optical, hamma2008adiabatic, trebst2007topological, emonts2020z3gauge, zohar2017z2gauge, burrello2021ladder}.
More specifically, in Sec.~\ref{sec:the_lattice_gauge_model} we introduce a pure $\Z_N$ gauge model on a (quasi-2D) ladder geometry, which includes both electric and magnetic degrees of freedom and admits superselection sectors, similarly to what happens in the Toric code.
To tackle the problem of gauge invariance, in Sec.~\ref{sec:duality_between_ladder_lgt_and_clock_models} we make use of a bond algebraic approach \cite{cobanera2011bond,nussinov2009bond} to introduce a duality transformation that allows for an exact mapping from the LGT on the ladder restricted to the gauge-invariant Hilbert space to a 1D $N$-clock model \cite{baxter1989clock,ortiz2012dualities,fendley2014parafermions,zhuang2015clock,sun2019phase} with a transversal field and a longitudinal field.
When periodic boundary conditions are enforced, the value of the latter turns out to depend on the super-selection sector of the ladder LGT, resulting in possible different phase diagrams for the different sectors, differently from what has been found previously with open boundary conditions \cite{burrello2021ladder}.

In Sec.~\ref{sec:numerical_investigations}, we resort to numerical analysis to study the phase diagram in the different sectors (labelled by $n=0, \dots, N-1$) for the $N=2,3$ and $4$ cases.
We first make use of exact diagonalization to determine: i) the presence of a deconfined-confined phase transition by calculating the value of the Wilson loops; ii) the ground state structure in the different phases.
We find that for all considered $N$, the model in the $n=0$ sector is always in the confined phase except for the deconfined point $\lambda=0$, where only the magnetic degrees of freedom are present in the Hamiltonian.
Instead, for $N$ even and $n=N/2$ we see a clear phase transition at about $\lambda=1$, while for $N=3$ and $n=1,2$ as well as for $N=4$ and $n=1,3$ we observe a cross-over region for $\lambda \lesssim 0.8$ followed by a phase transition to a double degenerate ground state.
A careful study of these cases requires longer chains and therefore is carried out via the DMRG algorithm.
Finally in Sec.~\ref{sec:conclusions_and_outlooks} we review our results and draw some conclusions.

\section{The lattice gauge model}
\label{sec:the_lattice_gauge_model}

Following the Hamiltonian approach of Kogut and Susskind \cite{kogut1975hamiltonian}, we consider a class of pure Abelian lattice gauge theories, with $\Z_N$ gauge group, on a \emph{ladder geometry}, which consists of a lattice $\mathbb{L}$ made of two parallels chains, the \emph{legs}, coupled to each other by \emph{rungs} to form square plaquettes.
On the ladder, each rung is identified by a coordinate $i=1,\dots,L$, where $L$ is the length of the ladder, and the two vertices on the rung are denoted with $i^{\uparrow}$ and $i^{\downarrow}$ in the upper and lower leg, respectively.
Links are denoted by $\ell$.
On the legs they are labelled as $\toplink_i$ (upper leg) or $\botlink_i$ (lower leg), while those on the rungs are labelled $\runglink_i$.

The gauge group degrees of freedom are defined on the links.
For a finite group like $\Z_N$, the notion of infinitesimal generators loses any meaning and we are led to directly consider, for each link $\ell \in \mathbb{L}$, a pair of conjugate operators, $U_{\ell} $ and $V_{\ell}$ which are unitary and defined by the algebraic relations \cite{schwinger1960unitary,notarnicola2015discrete, ercolessi2018znmodels}
\begin{equation}
    V_{\ell}  U_{\ell} = \omega U_{\ell} V_{\ell}, \qquad
    U^N_{\ell} = V^N_{\ell} = \One_{N}
    \label{eq:LGT_operators}
\end{equation}
with $\omega = e^{i\pqty{\frac{2\pi}{N}+\phi}}$, where the angle $\phi$ is arbitrary and corresponds to the physical situation in which on each link there is a background electric field \cite{ercolessi2018znmodels, magnifico2020realtimedynamics}.
In this manuscript we don't consider this situation and will set $\phi = 0$.
Also, these operators commute on different links.
This algebra admits a faithful finite-dimensional representation of dimension $N$ \cite{notarnicola2015discrete, weyl1950theory}.
To each link $\ell$, we associate an $N$-dimensional Hilbert space $\mathcal{H}_\ell$ generated by an orthonormal basis $\{\ket{v_{k,\ell}}\}$, with $k = 0, \dots, N-1$,  the \emph{electric basis}, that diagonalizes $V_\ell$:
\begin{equation}
    V_{\ell} \ket{v_{k,\ell}} = \omega^k \ket{v_{k,\ell}}.
\end{equation}
On this basis, $U_{\ell}$ acts as a shift operator,
\begin{equation}
    U_{\ell} \ket{v_{k, \ell} } = \ket{v_{k+1, \ell} },
\end{equation}
where $k+1$ is taken $\mathrm{mod}~N$.

\begin{figure}[t]
    \centering
    \includegraphics{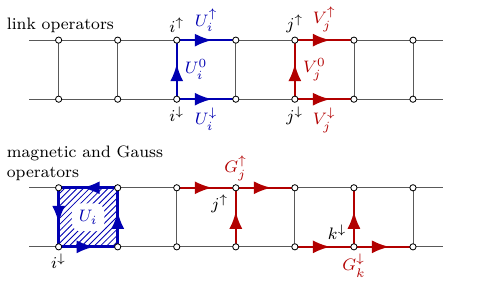}
    \caption{%
    Operators of the ladder $\Z_N$ LGT.
    \emph{Top panel:} operators $U_i^{\rho}$ and $V_j^{\rho}$ ($\rho = \uparrow, \downarrow, 0$) for each link;
    \emph{Bottom panel:} gauge-invariant magnetic operator $U_i$ and the Gauss operators $G_j^\uparrow$ and $G_k^\downarrow$.
}
    \label{fig:ladder_operators}
\end{figure}

As shown in the top panel of Fig.~\ref{fig:ladder_operators}, we use the symbols $V^0_i, \; U^0_i$ for the operators defined on the rung $i$, and  $V^{\rho}_i, \; U^{\rho}_i$ with $\rho = \uparrow, \downarrow$ for the operators on the horizontal links of the upper and lower leg to the right of the rung.
The links on the legs are oriented from left to right while those on the rungs from bottom to top.
To construct a LGT, in addition to the {\it electric field operators} $V$'s defined above, we need:
\vspace*{-5pt}
\begin{itemize}\itemsep0pt
    \item
        the \emph{magnetic operators}, which are defined on each plaquette to the right of the rung $i$ via the formula:
        \begin{equation}
            U_i = U^{\downarrow}_i \, U^0_{i+1} \, (U^{\uparrow}_i)^{\dagger} \, (U^0_i)^{\dagger};
        \end{equation}
    \item
        the  \emph{Gauss operators}, which are defined on each vertex $i^{\uparrow}, \; i^{\downarrow}$ of the lattice as:
        \begin{equation}
            G^{\uparrow}_i = \Vup_i (\Vup_{i-1})^{\dagger} (\Vh_i)^{\dagger}, \quad
            G^{\downarrow}_i = \Vdown_i \Vh_i (\Vdown_{i-1})^{\dagger}
            \label{eq:gauss_operators}
        \end{equation}
        and  implement local gauge transformations, by imposing that physical states should satisfy: $G^{\rho}_i \ket{\Psi_{\text{phys}}} = \ket{\Psi_{\text{phys}}} $ for $\rho = \uparrow, \downarrow$ and $\forall i$.
\end{itemize}
It is simple to verify that the $U_i$-operators commute with all Gauss operators, making them gauge invariant.
The operators defined so far are showed in the bottom panel of Fig.~\ref{fig:ladder_operators}.

The gauge-invariant Hamiltonian we use to build a $\Z_N$ LGT on the ladder can be written as:
\begin{equation}
    H_{\Z_N} = - \sum_{i}
    \bqty{
        U_i
        + \lambda \pqty{ \Vup_i + \Vdown_i + \Vh_i }
        + \text{h.c.}
    },
    \label{eq:hamiltoniana_ZN}
\end{equation}
with $\lambda > 0$, which is the relative strength between the electric and the magnetic fields.
One can choose to work with two separate couplings $\lambda_e$ and $\lambda_m$ for the electric and magnetic fields, respectively.
Nonetheless, we decided to use the ratio $\lambda = \lambda_e / \lambda_m$ for convenience and fix $\lambda_m = 1$, in order to work with just one parameter.
We use periodic boundary conditions on legs, which turns out to be an essential step for the duality map provided in the next section.

Similarly to what happens in the two dimensional Toric Code \cite{kitaev2003fault, tagliacozzo2011entanglement} (see in Appendix \ref{app:review_lgt}) the Hilbert space of physical states $\Hphys$ can be decomposed as a direct sum of superselection sectors $\Hphys^{(n)} $, where $n = 0, \dots, N-1$, that can be distinguished by means of the operators
\begin{equation}
    \overline{S} = \Vup_{i_0} \Vdown_{i_0}, \qquad
    \overline{W} = \prod_{i\in {\cal C}_0} U^{\downarrow}_{i},
    \label{eq:nonlocal_operators}
\end{equation}
where $i_0$ label the position of an arbitrary rung in the lattice, while $\mathcal{C}_0$ is any non-contractible loop around the ladder.
They satisfy the relations: $\overline{W} \, \overline{S} = \omega \overline{S} \,\overline{W}$.
Each physical state in a sector $\Hphys^{(n)}$ is an eigenstate of $\overline{S} $ with eigenvalue $\omega^n$, while $\overline{W}$ maps $\Hphys^{(n)}$ into $\Hphys^{(n+1)}$.

\section{Duality between ladder LGT and clock models}
\label{sec:duality_between_ladder_lgt_and_clock_models}

Clock models \cite{baxter1989clock, fendley2014parafermions, ortiz2012dualities} are a class of models that can be thought as a generalization of the quantum Ising model.
A $p$-state clock model on a chain has a local $p$-dimensional Hilbert space for each site $i=1,\dots,L$ and employs $p \times p$ unitary matrices $X_i$ and $Z_i$ that commute on different sites, while on the same site
\begin{equation}
    X_i Z_i = \omega Z_i X_i, \quad
    (X_i)^{p} = (Z_i)^p = \One_p,
    \label{eq:clock_operators}
\end{equation}
with $\omega = e^{i 2 \pi / p}$.
For example, one can choose a basis where the $Z_i$s are diagonal, i.e.~ $(Z_i)_{mn} = \delta_{m,n} \omega^{m}$ and $(X)_{mn} = \delta_{m,n+1}$ ($\mathrm{mod}~p$), with $m,n=0, \dots, p-1$.
The $p$-clock Hamiltonian is given by
\begin{equation}
    H_{p}(h)  = - \sum_{i} \left( Z_{i-1}^\dagger Z_{i} + h X_i + \text{h.c.} \right),
\end{equation}
where periodic boundary conditions are assumed and $h$ is the coupling of the transverse field.

We use the bond-algebraic approach to dualities \cite{cobanera2011bond}, to introduce a \emph{gauge reducing duality} mapping between the $\Z_N$ gauge model (with redundant degrees of freedom) on the ladder and an $N$-clock model on a single chain.
Similarly to what it can be done in 2D \cite{kogut1979introduction, fradkin1978order, cobanera2011bond}, we associate to each plaquette of the LGT a site of the chain of the clock model, in such a way that the gauge-invariant magnetic operator $U_i$ is mapped into the single-body operator $X_i$.
The duality of the two-dimensional gauge theories cannot be straightforwardly applied because here the links $\runglink$ have a different role when compared with the links $\toplink$ and $\botlink$, only the former being \emph{domain walls} between two plaquettes.
Also, the electric operators $\Vdown$/$\Vup$ on the top/bottom links $\toplink$/$\botlink$ have to be treated separately because they have different commutation relations with the plaquette operators $ U_i \equiv U^{\downarrow}_i \, U^0_{i+1} \, (U^{\uparrow}_i)^{\dagger} \, (U^0_i)^{\dagger}$:
\begin{equation}
    U_i \Vdown_i = \omega \Vdown_i U_i, \qquad
    U_i \Vup_i = \omega^{-1} \Vup_i U_i.
    \label{eq:comm_rel_ladder}
\end{equation}

\begin{figure}[t]
    \includegraphics{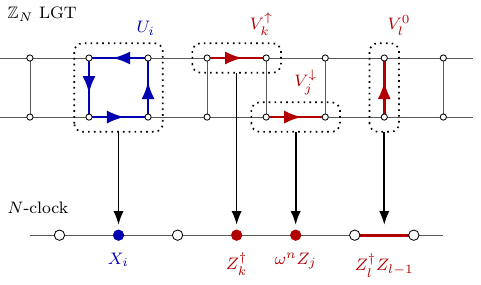}
    \caption{Visual representation of duality transformation from the $\Z_N$ ladder LGT to the $N$-clock model.}
    \label{fig:ladder_duality}
\end{figure}

The duality transformation is defined through the following steps.
First, the electric field on a vertical link $\runglink$ is mapped to $Z_{i-1}^\dagger Z_i$, as it is the result of the difference of the magnetic states of the two adjacent plaquettes.
This can be verified, since from the definition of the plaquette operator $U_{i}$ we get
\begin{equation*}
    V^0_i U_i = \omega^{-1} U_i V_i^0, \qquad
    V^0_i U_{i-1} = \omega U_{i-1} V_i^0,
\end{equation*}
therefore the maps
\begin{equation}
    U_i \mapsto X_i, \qquad
    V^0_i \mapsto Z_i^\dagger Z_{i-1},
    \label{eq:elec_h_and_plaq_op_map}
\end{equation}
conserve the commutation relations of $U_i$ and $V^0_i$.
Notice that, since from \eqref{eq:gauss_operators} we have
\begin{equation}
    \prod_{i} G_i^{\downarrow} \ket{\psi_{\text{phys}}} = \prod_{i} V_i^0 \ket{\psi_{\text{phys}}} = \ket{\psi_{\text{phys}}},
\end{equation}
we expect that, after the duality, the product of all $V_i^0$ is mapped to the identity, as it is from \eqref{eq:elec_h_and_plaq_op_map}.
This works as a check of consistency for the duality map.

Second, we consider $\Vup$ and $\Vdown$, that commute with $V^0$ while satisfy  relations \eqref{eq:comm_rel_ladder} with respect to $U_i$.
This allows us to assume:
\begin{equation}
    \Vdown_i \mapsto \coeffdown_i Z_i, \qquad
    \Vup_i \mapsto \coeffup_i Z_i^\dagger,
    \label{eq:elec_op_horiz_ladder_map}
\end{equation}
where $\coeffdown_i$ and $\coeffup_i$ are complex numbers, with $|\coeffdown_i|=|\coeffup_i|=~1$ to guarantee unitarity.
To further constraint the value of these coefficients, we have to impose that the Gauss constraints \eqref{eq:gauss_operators} become the identity: $G^\uparrow_i \mapsto \One$ and $G^\downarrow_i \mapsto \One$ for all $i$.
Since:
\begin{equation}
    \begin{split}
        G^\uparrow_i & \mapsto
        (\coeffup_i Z_i^\dagger) (\coeffup_{i-1} Z_{i-1}^\dagger) (Z_i^\dagger Z_{i-1})^\dagger
        = \coeffup_i (\coeffup_{i-1})^*, \\
        G^\downarrow_i & \mapsto
        (\coeffdown_i Z_i^\dagger) (Z_i^\dagger Z_{i-1}) (\coeffdown_{i-1} Z_{i-1}^\dagger)
        = \coeffdown_i (\coeffdown_{i-1})^*
    \end{split}
    \label{eq:gauss_law_map_ladder}
\end{equation}
we find that
\footnote{Thanks to \eqref{eq:gauss_law_map_ladder} we also know how to treat static matter.
Since it can be viewed as a violation of Gauss law, we just have to change the phases of $\coeffup_i$ and $\coeffdown_i$.}
\begin{equation}
    \coeffdown_i = \coeffdown, \qquad
    \coeffup_i = \coeffup, \qquad
    \forall i.
\end{equation}
for the following reason.
Given that all $\abs{c_i} = 1$, the condition $c_i (c_{i-1})^{\ast} = 1$ from \eqref{eq:gauss_law_map_ladder} is equivalent to $c_i = c_{i-1}$, which has to be true for all $i$.

Finally, since the superselection sectors are identified by the eigenvalue of the operator $\overline{S}$ in \eqref{eq:nonlocal_operators}, whose eigenvalues are simply $\omega^k$, for $k = 0, \dots, N-1$, we get
\begin{equation}
    \overline{S}
    \; \longmapsto \;
    ( \coeffup Z^\dagger_i ) ( \coeffdown Z_i )
    = \coeffup \coeffdown = \omega^k.
\end{equation}
This allows us to fix these coefficients as follows:
\begin{equation}
    \coeffup = 1, \qquad
    \coeffdown = \omega^k.
\end{equation}
We stress that this freedom of choice for the coefficients $\coeffup$ and $\coeffdown$ is due to the global $\Z_N$ of the system, not an effect of the already resolved gauge symmetry.

In summary, the duality mapping for the superselection sector $\omega^k$ of the $\Z_N$ LGT on a ladder reads as:
\begin{equation}
    \begin{aligned}
        U_i      & \; \longmapsto \; X_i, \quad &
        V^0_i    & \; \longmapsto \; Z^\dagger_{i-1} Z_i, \\
        \Vup_i   & \; \longmapsto \; Z_i^\dagger, \quad &
        \Vdown_i & \; \longmapsto \; \omega^k Z_i.
    \end{aligned}
\end{equation}
A sketch of this duality is given in Fig.~\ref{fig:ladder_duality}.
The transformed Hamiltonian is:
\begin{equation}
    H_{\text{lad}}(\lambda) \; \longmapsto \; \lambda H_{N}(\lambda^{-1})
\end{equation}
where
\begin{equation}
    H_{N} (\lambda^{-1}) = - \sum_{i}
    \pqty{
        Z_{i-1}^{\dagger} Z_i + \frac{1}{\lambda} X_i + (1 + \omega^k) Z_i
        + \text{h.c.}
    }.
    \label{eq:dual_ladder_hamiltonian}
\end{equation}
The novelty of \eqref{eq:dual_ladder_hamiltonian} is the appearance of a \emph{longitudinal field} term, with a coupling $(1 + \omega^n)$ that depends explicitly on the superselection sector $n$.
Notice that when $N$ is even, the longitudinal field is zero for $n = N/2$.
This simple fact makes it reasonable to think that different superselection sectors of the same ladder model can have drastically different phase diagrams.

Let us remark that the complex coupling $(1 + \omega^n)$ does not make the Hamiltonian (\ref{eq:dual_ladder_hamiltonian}) necessarily chiral \cite{fendley2012parafermions, whitsitt2018clock}.
In fact, one can get the real Hamiltonian
\begin{equation}
    H_N = H_p(1/\lambda) - 2 \cos \pqty{\frac{\pi n}{N}} \sum_{i} \pqty{Z_i + Z_i^{\dagger}}
    \label{eq:dual_ladder_hamiltonian_real}
\end{equation}
by absorbing the complex phase in the $Z_i$-operators, with the transformation $Z_i \mapsto w^{-n/2} Z_i$.
This transformation globally rotates the eigenvalues of the $Z_i$-operators, while preserving the algebra relations.
For $n$ even, this is just a permutation of the eigenvalues, meaning that it does not affect the Hamiltonian spectrum.
Instead, for $n$ odd, up to a reorder, the eigenvalues are shifted by an angle $\pi/N$, i.e.~half the phase of $\omega$.
The energy contribution of the extra term in \eqref{eq:dual_ladder_hamiltonian_real} depends on the real part of these eigenvalues and for $n$ odd we obtain that the lowest energy state is no longer unique.
In fact it is doubly degenerate.
This means that for $\lambda \to \infty$, where the extra term becomes dominant, we expect an ordered phase with a doubly degenerate ground state.
Finally, one can prove that the sectors $n$ and $N-n$ are equivalent
\footnote{For the sector $N-n$ we have that the overall factor $\cos(\pi(N-n)/N)$ is just $-\cos(\pi n/N)$.
    The minus sign can then be again absorbed into the $Z$'s operators.
This overall operation is equivalent to the mapping $Z \mapsto \omega^{-n/2} Z$ for the sector $N-n$.}.

\begin{figure}[t]
    \centering
    \includegraphics{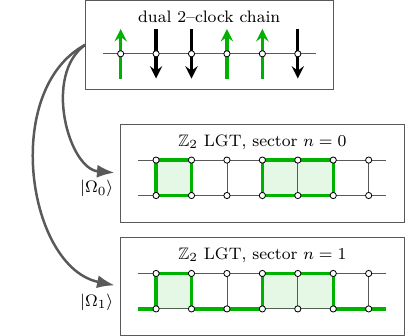}
    \caption{Duality between the states of a $2$--chain and the states of a $\Z_2$ ladder LGT in the different sectors $n=0$ (no non-contractible electric loop) and $n=1$ (one non-contractible loop around the ladder).
        In the sector $n=0$ it is evident that all the physical states contain closed electric loops.
        On the other hand, in the sector $n=1$ the physical states are all the possible deformations of the electric string that goes around the ladder.
    }
    \label{fig:z2_states}
\end{figure}

\section{Numerical investigations}
\label{sec:numerical_investigations}

We wish to investigate the presence of a \emph{deconfined-confined phase transition} (DCPT) for a given $\Z_N$ ladder LGT.
In a pure gauge theory, these phases can be detected
with the perimeter/area law for Wilson loops \cite{wilson1974confinement},
which can be expressed as the products of magnetic operators over a given region.
Unfortunately, in a ladder geometry there is not much difference between the area and the perimeter of a loop, since they both grow linearly in the size system $L$.

Nonetheless, we expect a phase transition by varying $\lambda$ \cite{trebst2007topological, hamma2008adiabatic, tagliacozzo2011entanglement} that can still be captured by an operator like $W_{\mathcal{R}}= \prod_{i \in \mathcal{R}} U_{i}$, the product of magnetic operators $U$'s over a (connected) region $\mathcal{R}$.
Indeed, when $\lambda=0$, the Hamiltonian \eqref{eq:hamiltoniana_ZN} is analogous to a Toric Code \cite{kitaev2003fault} which is known to be in a deconfined phase, where the (topologically distinct) ground states are obtained as uniform superpositions of the gauge-invariant states, i.e.~closed electric loops.
On these ground states $\ev{W_{\mathcal{R}}} = 1$, hence
a value $\ev{W_{\mathcal{R}}} \approx 1$ signals a deconfined phase.
On the other hand, when $\lambda \rightarrow \infty$, the electric loops are suppressed, hence
$\ev{W_{\mathcal{R}}} \approx 0$, signalling a confined phase.

In the dual clock model picture, the Wilson loop translates to a disorder operator \cite{fradkin1978order}, which means that a deconfined phase can be thought of as a paramagnetic (or disordered) phase, while the confined phase is like a ferromagnetic (or ordered) phase.
Moreover, the longitudinal field breaks the $N$-fold symmetry of the ferromagnetic phase into a one-fold or two-fold degeneracy, depending on the superselection sector.

We first start by studying the $\Z_N$ LGT on a ladder numerically through \emph{exact diagonalization}, by evaluating the half-ladder Wilson loop, i.e.~
\begin{equation}
    W = U_1 U_2 \cdots U_{L/2},
    \label{eq:half_ladder_wilson}
\end{equation}
on the ground state while working in the restricted physical Hilbert space $\Hphys^{(n)}$ ($n=0,\dots,N-1$), which has dimension $N^L$, much smaller than $N^{3L}$ (the dimension of the total Hilbert space).
Then, we procede to analyze some region of interest by means of \emph{DMRG} \cite{schollwock2011dmrg}, but on the corresponding dual clock model.

In the following, we will present the results for $N=2,3$ and $4$, but before doing so we will discuss the implementation of the physical Hilbert space for the exact diagonalization.

\begin{figure*}[t]
    \centering
    \subfloat[$N=2$]{%
        \includegraphics{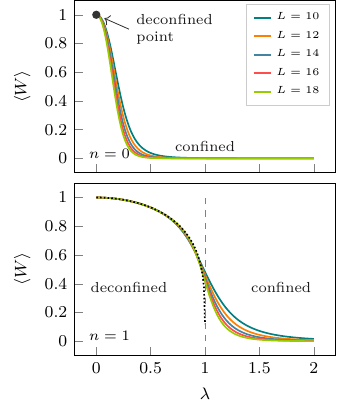}%
        \label{fig:wilson_loops_Z2}%
    } ~%
    \subfloat[$N=3$]{%
        \includegraphics{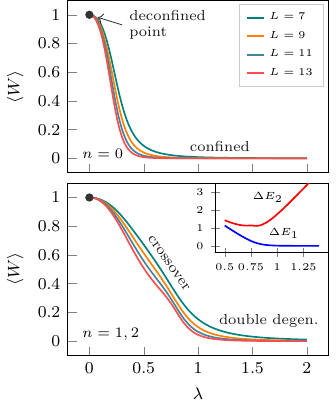}%
        \label{fig:wilson_loops_Z3}%
    } ~%
    \subfloat[$N=4$]{%
        \includegraphics{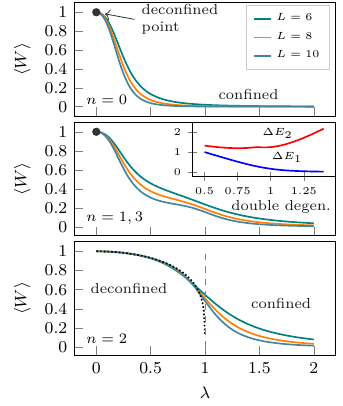}%
        \label{fig:wilson_loops_Z4}%
    }%

    \vspace{-5pt}
    \caption{
        Expectation values of the half-ladder Wilson loop \eqref{eq:half_ladder_wilson} of the ground state of the $\Z_N$ ladder LGT.
        They have been computed using exact diagonalization in each superselection sector for different lattice sizes, using 100 values of $\lambda$ in the range $0 \leq \lambda \leq 2$.
        \\
        \textbf{(a)} Case $N=2$ for sizes $L = 10, 12, \dots, 18$ and sectors $n=0$ (\emph{top}) and $n=1$ (\emph{bottom}).
        Only for $n=1$ we have a clear phase transition for $\lambda \simeq 1$, while $n=0$ is always confined for $\lambda \neq 0$.
        The limit for large $L$ of the Wilson loop \eqref{eq:wilson_loop_limit} is shown with a dotted line. \\
        \textbf{(b)} Case $N=3$ for sizes $L = 7,9,11,13$ and sectors $n=0$ (\emph{top}) and $n=1,2$ (\emph{bottom}, which are equivalent).
        In the latter we see the appearance of a crossover region and a double degenerate ordered phase.
        \emph{Inset:} energy gaps $\Delta E_i = E_i - E_0$ for $i=1,2$ and size $L = 13$, as a function of the coupling $\lambda$, in the sectors $n=1,2$, showing the emergence of a double-degenerate ground state for $\lambda > 1$. \\
        \textbf{(c)} Case $N=4$ for sizes $L=6,8,10$ and sectors $n=0$ (\emph{top}), $n=1,3$ (\emph{middle}, which are equivalent) and $n=2$ (\emph{bottom}).
        Only the sector $n = 2$ has a clear deconfined-confined phase transition, as expected from the duality with the $4$-clock model, while for $n=1,3$ the situation is similar to $N=3$ and $n=1,2$.
        The limit for large $L$ of the Wilson loop for $n=2$ is the same as \eqref{eq:wilson_loop_limit}, shown with a dotted line. This is because a 4-clock model is equivalent to two Ising models \cite{ortiz2012dualities}. \\
        \emph{Inset:} energy gaps $\Delta E_i = E_i - E_0$ for $i=1,2$ and size $L=10$. A situation similar to $N=3$ and $n=1,2$ arises.
    }
    \label{fig:wilson_loops}
\end{figure*}

\subsection{Implementation of the Gauss Law}
\label{sub:implementation_of_the_gauss_law}

When considering a LGT, one would like to work within the physical subspace, which is obtained by imposing Gauss law at every site.
A straightforward but inefficient method, in which one generates all the possible states and then filters out all the states that violate Gauss law, is not efficient, even for moderately small lattices.
To better exemplify this, consider a $\Z_2$ theory on a $L \times L$ periodic lattice, which have $L^2$ sites and $2L^2$ links, and only $2^{L^2}$ physical states.
There are therefore $2^{2 L^2}$ possible states and for each one up to $L^2$ checks (one per site) has to be performed.
As a result, the construction of the physical Hilbert space involves $O(L^2 2^{2 L ^2})$ operations in a search space of $2^{2 L^2}$ objects for finding only $2^{L^2}$ elements.
Here, we exploit the gauge-reducing duality map described in Sec.~\ref{sec:duality_between_ladder_lgt_and_clock_models} for the ladder one, to devise an algorithmic procedure that generates physical configurations starting from the states of the dual clock model.
This is not a search or pattern-matching algorithm and gives a \emph{major speedup} with respect to the direct method just described.
A similar approach has been used in \cite{kaplan2020gauss} for a $U(1)$ gauge theory, where however an overcomplete basis is found.

Given a $\Z_N$ LGT on a lattice of size $L \times L$, we consider the dual $N$-clock model on a similar lattice with $A = L^2$ sites.
A basis for the Hilbert space of the clock-model is the set of states $\ket{ \{s_i\} } \equiv \ket{s_0 s_1 \cdots s_{A-1}}$, with $s_i = 0, \dots, N-1$.
The corresponding gauge-invariant state in each superselection sector $\Hphys^{(n,m)}$ of the Hilbert space of the dual LGT model is given by:
\begin{equation}
    \ket{\{s_i\} } \; \longmapsto \;
    \prod_{i=0}^{A-1} U_i^{s_i} \ket{\Omega_{(n,m)}},
\end{equation}
where $U_i$ is the plaquette operator on the $i$-th plaquette and $\ket{\Omega_{(m,n)}}$ is the ``Fock vacuum'' of the $\Hphys^{(n,m)}$ subspace.
Moreover, the ``Fock vacuums'' $\ket{\Omega_{(n,m)}}$ can be obtained as:
\begin{equation}
    \ket{\Omega_{(n,m)}} = (\overline{W}_1)^n (\overline{W}_2)^m \ket{\Omega_{(0,0)}},
\end{equation}
where $\ket{\Omega_{(0,0)}}$ is the vacuum in the $(0,0)$-sector, i.e.~the state $\ket{000 \cdots 0}$.
For more details see Appendix \ref{app:review_lgt}.
In the case of a ladder geometry, where $m$ is always zero, we shorten the notation of the vacuum states to $\ket{\Omega_n}$.
Fig.~\ref{fig:z2_states} show some examples of physical states in different superselection sectors.

Let us quantify the obtained speedup with this method.
In the case of a $\Z_2$ theory on a square lattice $L \times L$ there are $2^{L^2}$ possible clock configurations.
For each configuration, there are at most $L^2$ magnetic fluxes to apply.
This translates into $O(L^2 2^{L^2})$ operations: notice that the exponent does not contains the factor 2 which is present in the direct but inefficient method, thus reducing the number of operations by an order of $O(2^{L^2})$.
The procedure is generalizable for any $\Z_N$.
This algorithm has been developed independently but similar techniques for different models can be found, for example, in \cite{kaplan2020gauss}.

\subsection{Exact diagonalization for \texorpdfstring{$N=2$}{N=2}}
\label{sub:results_N_2}%

As a warm up, we consider the $\Z_2$ ladder LGT, with lengths $L=10,12,\dots,18$.
This model is equivalent to a $p=2$ clock model, which is just the quantum Ising chain, with only two superselection sectors for $n=0$ and $n=1$.
When $n=1$, the Hamiltonian contains only the transverse filed and is integrable \cite{baxter1982exactlysm}.
Thus, we expect a critical point for $\lambda \simeq 1$, which will be a DCPT in the gauge model language.
This is seen in the behaviour of the half-ladder Wilson loop, as shown in the lower panel of Fig.~\ref{fig:wilson_loops_Z2}.
For $n=0$, both the transverse and longitudinal fields are present, the model is no longer integrable \cite{banuls2011thermalization, kormos2017confinement, pomponio2022bloch} and we expect to always see a confined phase, except for $\lambda = 0$.
This is indeed confirmed by the behaviour of the half-ladder Wilson loop shown in the upper panel of Fig.~\ref{fig:wilson_loops_Z2}.

Regarding the Wilson loop in the sector $n=1$, it is possible to compute its behaviour for large $L$ using a Kramer-Wannier duality.
Consider $\lambda H_N(\lambda^{-1})$ in \eqref{eq:dual_ladder_hamiltonian} for $N=2$ and $n=1$.
This model is self-dual under the following Kramer-Wannier transformation:
\begin{equation}
    X_i \mapsto Z_i Z_{i+1} \qquad
    Z_{i-1} Z_i  \mapsto X_i.
    \label{eq:kramers_wannier}
\end{equation}
Note that $Z = Z^{\dagger}$ and $X = X^{\dagger}$ for $N=2$.
With the above map we obtain
\begin{equation*}
    \lambda H_{N=2} (\lambda^{-1}) \mapsto H_{N=2}(\lambda),
\end{equation*}
where now the coupling $\lambda$ acts as a transverse field, hence for $\lambda < 1$ we have a ferromagnetic phase signaled by the magnetization $M = \sum_{i} \ev{Z_i} / L$.

Furthermore, the half-ladder Wilson loop \eqref{eq:half_ladder_wilson}  after the gauge-reduction can be written as $W = X_1 \cdots X_{L/2}$, which under the map \eqref{eq:kramers_wannier} becomes
\begin{equation}
    W = Z_1 Z_{1 + L/2}.
\end{equation}
Such correlator in the limit of large $L$ reduces to the magnetization squared:
\begin{equation*}
    \ev{W} \sim \ev{Z}^2 = M^2 \quad \text{for $L \gg 1$}.
\end{equation*}
Thanks to the Onsager formula, we have a closed analytical expression for the magnetization,
which in the one-dimensional quantum Ising model translates to \cite{pfeuty1970ising}
\begin{equation*}
    M \sim \left(1 - \lambda^2 \right)^{1/8},
\end{equation*}
therefore in the limit of large $L$ we have
\begin{equation}
    \ev{W} \sim (1 - \lambda^2)^{1/4}.
    \label{eq:wilson_loop_limit}
\end{equation}

This curve is shown in the lower panel of Fig.~\ref{fig:wilson_loops_Z2} (dotted line) to compare it with the numerical results.
Close to the phase transitions ($\lambda = 1$) the numerical data suffers from strong finite-size corrections, while they are in good agreement with the theoretical result in the region where $\lambda$ is small and we are deep inside the deconfined phase.

We can further characterize the phases of the two sectors by looking at the structure of the ground state, for $\lambda<1$ and $\lambda>1$, which is possible thanks to the exact diagonalization (see Appendix \ref{app:ground_state_distribution}).
In particular, in the deconfined phase of the sector $n=1$, the ground state is a superposition of the deformations of the non-contractible electric string that makes the $n=1$ vacuum $\ket{\Omega_1}$.
For this reason, this phase can be thought as a \emph{kink condensate} \cite{fradkin1978order} (which is equivalent to a paramagnetic phase), where each kink corresponds to a deformation of the string.
Instead, for $\lambda > 1$, where we have confinement (as in the $n=0$ sector), the ground state is essentially a product state, akin to a ferromagnetic state.

\subsection{Exact diagonalization for \texorpdfstring{$N=3$}{N=3}}%
\label{sub:results_N_3}%

The $\Z_3$ LGT is studied for lengths $L=7,9,11$ and $13$.
This model can be mapped to a $3$-clock model, which is equivalent to a $3$-state quantum Potts model, with a longitudinal field presents in all sectors, as one can see from \eqref{eq:dual_ladder_hamiltonian_real}.
This field is expected to disrupt any ordered state in the $X$-basis, hence any deconfined phase in the gauge model.
Thus it is not possible to observe a phase transition, as confirmed by the behaviour of the half-ladder Wilson loops in Fig.~\ref{fig:wilson_loops_Z3}.
As expected, all the sectors present a deconfined point at $\lambda = 0$.

In the case $n=0$, for $\lambda > 0$ we recognize a quick transition to a confined phase, similar to what happens in \cite{burrello2021ladder}.
While for $n=1$ and $2$ (which are equivalent), the model exhibits a smoother \emph{crossover} to an ordered phase characterized by a doubly-degenerate ground state, for $\lambda > 1$.
In the crossover region the Wilson loops decrease much slower with respect to the $n=0$ sector.
This could point to a new phenomenology that appears in the sectors $n=1,2$.
A more detailed analysis of this crossover region can be found in Sec.~\ref{sub:crossover_region}.
Notice that, as discussed above, the presence of the ``skew'' longitudinal field breaks the three-fold degeneracy, expected in an ordered phase of the $3$-clock model, into a two-fold degeneracy only.

\begin{figure*}[t]
    \centering
    \subfloat[Energy gaps]{%
        \vspace*{5pt}  %
        \includegraphics{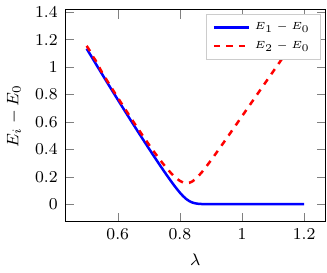}%
        \label{fig:dual_energy_gaps_N3}%
    }
    ~~
    \subfloat[Wilson loop]{%
        \includegraphics{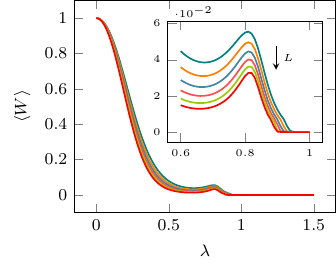}%
        \label{fig:dual_wilson_loop_N3}%
    }
    ~~
    \subfloat[Magnetization]{%
        \includegraphics{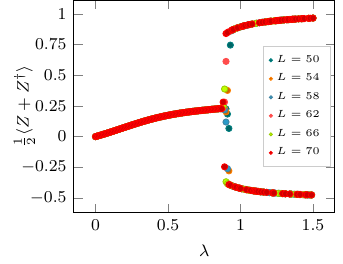}%
        \label{fig:dual_magnetization_N3}%
    }
    \caption{Numerical analysis of the dual clock model for $N=3$ and $n=1$ with DMRG.
        The expectation values have been computed on the ground state in the bulk of the chain, by which we mean the region from $L/4$ to $3L/4$, in order to avoid finite size effects.
        The shown quantities have been calculated for 150 values of $\lambda$ in the region $0 \leq \lambda \leq 1.5$.\\
    (a) Energy gaps of the first two excited states at size $L=70$; we can see that the large $\lambda$ region is doubly degenerate. \\
    (b) Equivalent of the half-ladder Wilson loop, which is the expectation value of the disorder operator $W = X_{L/4} \cdots X_{3L/4}$.
    The lattice sizes are $L=50,54,\dots,70$, following the arrow direction.
    In the inset a focus on the region $0.6 \leq \lambda \leq 1$ is shown, where bumps are present.
    In the region $\lambda \lesssim 0.6$ the Wilson loop quickly decreases to zero, while for $\lambda > 1$ it is vanishing.
    Notice, however, that the heights of the bumps is still small when compared to the deconfined phase.
    \\
    (c) Magnetization, by which we mean the expectation value of $\frac{1}{2} (Z + Z^{\dagger})$ averaged over the lattice sites in the bulk. Notice we have a bifurcation of the order parameter for $0.87 \lesssim \lambda \lesssim 0.9$.
    }%
    \label{fig:dual_clock_N3}

\end{figure*}

\subsection{Exact diagonalization for \texorpdfstring{$N=4$}{N=4}}%
\label{sub:results_N_4}%
The $\Z_4$ ladder LGT have four superselection sectors.
The behaviour of half-ladder Wilson loops as function of $\lambda$ is shown in Fig.~\ref{fig:wilson_loops_Z4}.
As in the previous models, for $n=0$ we see a deconfined point at $\lambda = 0$, followed by a sharp transition to a confined phase.
The sector $n=2$, which has no longitudinal field, is the only one to present a clear DCPT for $\lambda \approx 1$, as it is expected from the fact that the $4$-clock model is equivalent to two decoupled Ising chains \cite{ortiz2012dualities}.

In the two equivalent sectors $n=1$ and $3$, where the longitudinal field coupling is complex, the Wilson loop shows a peculiar behaviour, at least for the largest size ($L=10$) of the chain: it decreases fast as soon $\lambda > 0$, to stabilize to a finite value in the region $0.5 \lesssim \lambda \lesssim 1$, before decreasing to zero.
As for the $N=4$ case, we present a deeper analysis of this situation in Sec.~\ref{sub:crossover_region}.
For $\lambda \gtrsim 1$, the system enters a confined phase with a double degenerate ground state, as for the $\Z_3$ model.

\subsection{DMRG analysis of the crossover region}
\label{sub:crossover_region}

In this section we further analyze the crossover regions and the possible transition point that appears in the $N=3$ and $N=4$ cases.
In particular, we focus on the following cases: (i) $N = 3$ sector $n=1$; (ii) $N=4$ sector $n=1$.
We chose to do so by directly studying the dual clock model with DMRG techniques, which allow us to access much larger lattice sizes.
The results obtained with exact diagonalization on the ladder gauge theory all confirm the duality discussed in Sec.~\ref{sec:duality_between_ladder_lgt_and_clock_models}, henceforth we no longer feel the need to study the gauge model instead.
For the DMRG simulations we used open boundary conditions, a maximum bond dimension $\chi = 800$, and a cutoff of $10^{-10}$.
Additionally, in order to avoid finite-size effects, all the expectation values have been computed on the ground state in the bulk of the system, i.e. in the region from $L/4$ to $3L/4$.

\begin{figure*}[t]
    \centering
    \subfloat[Energy gaps]{%
        \vspace*{5pt}  %
        \includegraphics{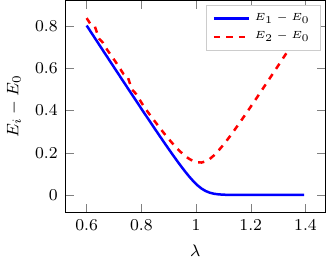}%
        \label{fig:dual_energy_gaps_N4}%
    }
    ~~
    \subfloat[Wilson loop]{%
        \includegraphics{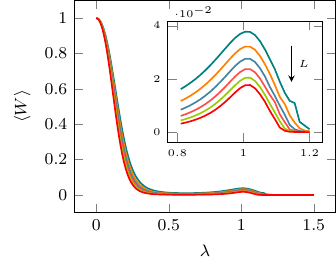}%
        \label{fig:dual_wilson_loop_N4}%
    }
    ~~
    \subfloat[Magnetization]{%
        \includegraphics{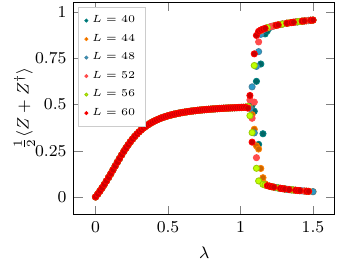}%
        \label{fig:dual_magnetization_N4}%
    }
    \caption{Numerical analysis of the dual clock model for $N=4$ and $n=1$ with DMRG.
        The expectation values have been computed on the ground state in the bulk of the chain, by which we mean the region from $L/4$ to $3L/4$, in order to avoid finite size effects.
        The shown quantities have been computed for 100 values of $\lambda$ in the region $0 \leq \lambda \leq 1.5$.\\
    (a) Energy gaps of the first two excited states at size $L=60$; we can see that the large $\lambda$ region is doubly degenerate. \\
    (b) Equivalent of the half-ladder Wilson loop, which is the expectation value of the disorder operator $W = X_{L/4} \cdots X_{3L/4}$.
    The lattice sizes are $L=50,54,\dots,70$, following the arrow direction.
    In the inset a focus on the region $0.8 \leq \lambda \leq 1.2$ is shown, where bumps are present.
    The only clear difference with the $N=3$ case is the presence of a more consistent plateau, for $\lambda \lesssim 0.8$, where the Wilson loop vanishes.\\
    (c) Magnetization, by which we mean the expectation value of $\frac{1}{2} (Z + Z^{\dagger})$ averaged over the lattice sites in the bulk.
    Here the bifurcation point has moved to $\lambda \approx 1$.
    }
\end{figure*}

First of all, we confirm that for large $\lambda$ we have a doubly degenerate ordered phase by looking at Figs.~\ref{fig:dual_energy_gaps_N3} and \ref{fig:dual_energy_gaps_N4}.
They show, respectively for $N=3$ and $N=4$, the energy gaps of the first and second excited levels, with respect to the ground state energy $E_0$.
In both cases, $E_1 - E_0$ goes to zero for large $\lambda$.

Furthermore, in the crossover region of Figs.~\ref{fig:wilson_loops_Z3} and \ref{fig:wilson_loops_Z4} we notice a small bump, even though the lattice size is quite small.
These bumps become more evident for larger $L$, as it can be seen in Figs.~\ref{fig:dual_wilson_loop_N3} and \ref{fig:dual_wilson_loop_N4}.
But some rough finite-size scaling shows that the maxima of the bumps do not reach a positive non-zero value in the limit $L \to \infty$.
In order to reach a more definite answer, simulations of much larger scales is necessary, as the results presented here are just qualitative.
We speculate that the origin of these bumps is due to the transition to the doubly degenerate ordered phase.

Since we have an ordered phase for large $\lambda$, we expect a non-zero value of the magnetization.
Therefore, we computed the expectation value of $ (Z + Z^{\dagger})/2$ as a function of $\lambda$ and the result is shown in Fig.~\ref{fig:dual_magnetization_N3} for $N=3$, and in Fig.~\ref{fig:dual_magnetization_N4} for $N=4$.
The remarkable feature of this magnetization is that a bifurcation arises in both cases.
For $N=3$ it appears around $0.87 \lesssim \lambda \lesssim 0.92$, while for $N=4$ it is around $1.0 \lesssim \lambda \lesssim 1.1$.
The origin of this bifurcation is explained by the doubly degenerate ground states, each with a different value of the magnetization.
Due to the lack of conserved quantum numbers, the DMRG algorithm can end up in any of the two cases randomly for each run.

The two degenerate ground states can be easily described.
They have to be ferromagnetic states, due to the minus sign in front of $Z^{\dagger} Z$ in \eqref{eq:dual_ladder_hamiltonian}, which means all the sites have to be aligned along the same eigenvector in the $Z$-basis.
Moreover, the two possible alignments are given by the highest weight eigenvectors of the longitudinal field $(1 + \omega^n)Z + (1 + \omega^{-n})Z^{\dagger}$.
Then, the corresponding magnetization is just given by their eigenvalues with respect to $(Z + Z^{\dagger})/2$.
Let $\ket{s}$ be a ket in the $Z$-basis with eigenvalue $\omega^s$.
With some simple algebra we find the following:
\begin{enumerate}\itemsep0pt
    \item For $N=3$ and $n=1$ the highest weight eigenvectors are $\ket{0}$ and $\ket{2}$, with magnetization $1$ and $-1/2$ respectively.
    \item For $N=4$ and $n=1$ instead we have $\ket{0}$ and $\ket{3}$, with magnetization $1$ and $0$ respectively.
\end{enumerate}
These values are confirmed by Figs.~\ref{fig:dual_magnetization_N3} and \ref{fig:dual_magnetization_N4} in the limit of large $\lambda$.

The ground states in the large $\lambda$ limit have a clear interpretation in the gauge model.
Consider the case $N=3$ and sector $n=1$.
Following the duality Sec.~\ref{sec:duality_between_ladder_lgt_and_clock_models} and the procedure in Sec.~\ref{sub:implementation_of_the_gauss_law}, one can see that the clock state $\ket{0 \cdots 0}$ corresponds to the situation where all the links in the lower leg are in the $\ket{1}$ state, while the rest is in $\ket{0}$.
Conversely, the clock state $\ket{2 \cdots 2}$ corresponds to the configuration where the upper leg is in the $\ket{2}$ state.
Similar pictures can be drawn for $N=4$ and $n=1$, where it is either the lower leg in the $\ket{1}$ state or the upper leg in the $\ket{3}$ state.

A surprising feature of the crossover region is the behaviour of the magnetization.
Consider the case $N=3$.
If the phase $\lambda \lesssim 0.87$ is indeed paramagnetic, one would expect a vanishing magnetization.
Instead, we find that it follows a profile where it slowly grows, until it reaches a maximum close to $0.25$ around $\lambda \approx 0.87$, just before the bifurcation.
We exclude the possibility of finite size effects, because the magnetization is computed in the bulk and its behaviour is independent of the chain size, as it can be seen in Fig.~\ref{fig:dual_magnetization_N3}.
This maximum is close to the average of the two magnetization in the large $\lambda$ limit, which suggest that ground state may be close to a superposition $(\ket{0 \cdots 0} + \ket{2 \cdots 2})/ \sqrt{2}$.
In the region close to the transition point ($\lambda \sim 1$) we observe that the numerical data for the magnetization are very scattered while Wilson loops show a non-zero bump in the expectation value.
We interpret this as a signal that, in this region, the numerics is strongly affected by the presence of low-energy disordered states, that are instead suppressed deeper in the confined region.

The same argument can be repeated for $N=4$, the only noticeable difference is a clear plateau of vanishing Wilson loop before the bifurcation.
This means that the ground state is much closer to the superposition $(\ket{0 \cdots 0} + \ket{3 \cdots 3}) / \sqrt{2}$ in this case.

\section{Conclusions and outlooks}
\label{sec:conclusions_and_outlooks}

In this work, we proposed an exact gauge reducing duality transformation that maps the $\mathbb{Z}_N$ lattice gauge theory on a ladder onto a 1D $N-$clock model in a transversal field, coupled to a possibly complex longitudinal field which depends on the superselection sector.

This map allowed us to perform numerical simulations with an exact diagonalization algorithm with sizes up to $L=18, 13, 10$ for $N=2,3,4$ respectively.
To study the phases of the model and a possible DCPT point, we calculated the Wilson loops in the different superselection sectors.
For $N$ even and $n=N/2$ we obtain a DCPT point;
for $n=0$ and any $N$ we are always in a confined phase when $\lambda \neq 0$;
while we find an unusual behaviour in the other cases ($N = 3$ with $n=1,2$ and $N=4$ with $n=1,3$).
In particular, the case of $n=1$ for $N=3,4$ have been further analyzed using DMRG techniques and the results suggest the emergence of a phase that cannot be properly called paramagnetic.

We have shown that the phase diagram of these gauge models depends heavily on the superselection sectors.
Such sectors exist only with periodic boundary conditions but one can obtain a similar setup with open boundary conditions instead.
It is sufficient to fix the electric flux at the ends of the ladder, which will be effectively equivalent at looking at the bulk of the periodic lattice.
Therefore, boundary conditions play a key role in the phenomenology of these models.

The results presented here regarding the odd sectors are just qualitative and they deserve a proper analysis, that we plan to do in the future.
Another possible direction for future work is the inclusion of matter (static or dynamical) in these gauge models, and how they affect the correspondence with quantum clock models.

\begin{acknowledgments}
The numerical analysis have been performed with the QuSpin library \cite{weinberg2017quspin, weinberg2019quspin} for the exact diagonalization, and the ITensor library \cite{itensor, itensor-r0.3} for the DMRG simulations.
We thank M.~Burrello and O.~Pomponio for useful discussions.
This research is partially supported by INFN through the project “QUANTUM”, the project “SFT" and the project “QuantHEP" of the QuantERA ERA-NET Co-fund in Quantum Technologies (GA No. 731473).
\end{acknowledgments}

\appendix

\section{Review of two-dimensional LGTs and the Toric Code}
\label{app:review_lgt}

In this appendix we review some aspects of LGTs in two dimensions.
For a more general review we suggest \cite{kitaev2003fault, tagliacozzo2011entanglement}.

For a discrete group like $\mathbb Z_N$, the notion of infinitesimal generators loses any meaning and we are led to directly consider, for each link $\ell \in \mathbb L$, two unitary operators
$V_\ell, \, U_\ell$, such that \cite{schwinger1960unitary, schwinger2001symbolism}
\begin{equation}
    V_\ell U_\ell V_\ell^{\dagger}=e^{2\pi i/N}U_\ell, \qquad
    U_\ell^N=\One_N, \qquad
    V_\ell^N=\One_N.
    \label{eq:schwinger_weyl_algebra}
\end{equation}
while on different links they commute.
Thus, by representing $\Z_N$ with the set of the $N$ roots of unity $e^{i 2 \pi k/N}$\, ($k=1, \cdots, N$), commonly referred to as the discretized circle,
we see that $V$ plays the role of a ``position operator'' on the discretized circle, while $U$ that of a ``momentum operator''.

These algebraic relations admit a faithful finite-dimensional representation of dimension $N$ \cite{weyl1950theory}, for any integer $N$, which is obtained as follows.
To each link $\ell$, we can associate an $N$-dimensional Hilbert space $\mathcal{H}_\ell$ generated by an orthonormal basis $\{\ket{v_{k,\ell}}\}$ ($k=1, \dots,N$), called the \emph{electric basis},
that diagonalizes $V_\ell$.
With this choice, we can promptly write the actions of $U_\ell$ and $V_\ell$:
\begin{equation}
    \begin{split}
        U\ket{v_{k,\ell}}           = \ket{v_{k+1,\ell}}, & \qquad
        U\ket{v_{N,\ell}}           = \ket{v_{1,\ell}}\\
        U^{\dagger}\ket{v_{k,\ell}} = \ket{v_{k-1,\ell}}, & \qquad
        U^\dagger\ket{v_{1,\ell}}   = \ket{v_{N,\ell}}\\
        V\ket{v_{k,\ell}}           = \omega^k \ket{v_{k,\ell}}, & \qquad
        V^\dagger\ket{v_{k,\ell}}   = \omega^{-k} \ket{v_{k,\ell}},
    \end{split}
    \label{eq:elect_basis_op_action}
\end{equation}
where $\omega = e^{2 \pi i / N}$ and $k = 0, \dots, N-1$.
We choose to work in this particular basis and the various $k$ can be interpreted as the quantized values of the electric field on the links.



\subsection{Gauge invariance and physical states}%
\label{sub:gauge_invariance}

\begin{figure}[t]
    \centering
    \includegraphics{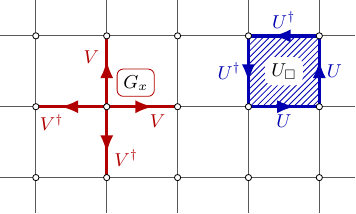}
    \caption{Pictorial representation of the Gauss operators $G_x$ in \eqref{eq:gauss_operator} (\emph{left}) and plaquette operator $U_{\square}$ in \eqref{eq:plaq_operator} (\emph{right}).}
    \label{fig:star_plaq_operators}
\end{figure}

Gauge transformations act on vector potentials while preserving the electric field.
In the case of a discrete symmetry, a gauge transformation at a site $x \in \L$ is a product of $V$'s (and $V^\dagger$'s) defined on the links which comes out (and enters) the vertex.
More specifically, for a two dimensional lattice,
if the link $\ell$ at site $x$ is oriented in the positive direction, i.e.~either $(x, +\hat{1})$ or $(x, +\hat{2})$, then $V$ is used, otherwise $V^\dagger$.
Thus,
the single local gauge transformation at the site $x$ is enforced by the operator:
\begin{equation}
    G_x =
    V_{(x, \hat{1})}^{\phantom{\dagger}}
    V_{(x, \hat{2})}^{\phantom{\dagger}}
    V^\dagger_{(x, -\hat{1})}
    V^\dagger_{(x, -\hat{2})},
    \label{eq:gauss_operator}
\end{equation}
as shown in the left part of in Fig.~\ref{fig:star_plaq_operators}.

The total Hilbert space $\mathcal{H}^{\text{tot}}$ is given by the $\otimes_{\ell} \mathcal{H}_{\ell}$.
A state of the whole lattice $\ket{\Psi_{\text{ph}}} \in \mathcal{H}^{\text{tot}}$ is said to be \emph{physical} if it is a \emph{gauge-invariant state}:
\begin{equation}
    G_x \ket{\Psi_{\text{ph}}} = \ket{\Psi_{\text{ph}}}, \qquad \forall x \in \L
    \label{eq:gauss_law}
\end{equation}
This condition can be translated into a constraint on the eigenvalues $v_{(x, \pm \hat{i})}= \omega^{k_{(x, \pm \hat{i})} }$ of the operators $V_\ell$ on the links $\ell = (x, \pm \hat{i})$ of the vertex $x$:
\begin{equation}
    v_{(x, \hat{1})}^{\phantom{\ast}}
    v_{(x, \hat{2})}^{\phantom{\ast}}
    v_{(x, -\hat{1})}^\ast
    v_{(x, -\hat{2})}^\ast = 1,
\end{equation}
or, because of (\ref{eq:elect_basis_op_action}):
\begin{equation}
    \sum_{i=1,2} \pqty{ k_{(x, \hat{i})} - k_{(x, -\hat{i})} } = 0 \quad \text{mod $N$}.
    \label{eq:gauss_law_elec_eigvals}
\end{equation}
Given the fact that the $k$'s in \eqref{eq:schwinger_weyl_algebra} represent the values of the electric field, one can see that \eqref{eq:gauss_law_elec_eigvals} can be interpreted as a discretized version of the Gauss law $\nabla \cdot \vec{E} = 0$ in two dimensions,
for a pure gauge theory where there are no electric charges.

One can see that the electric operators $V_{\ell}$ and the plaquette operators
\begin{equation}
    U_{\square} =
    U_{(x, \hat{1})}
    U_{(x + \hat{1}, \hat{2})}
    U_{(x + \hat{1} + \hat{2}, -\hat{1})}^\dagger
    U_{(x + \hat{2}, -\hat{2})}^\dagger.
    \label{eq:plaq_operator}
\end{equation}
are gauge invariant.
Then, the Hamiltonian of the model can be written as
\begin{equation}
    H_{\Z_N}(\lambda) = - \sum_{\square} U_{\square} - \lambda \sum_{\ell} V_{\ell} + \text{h.c.}
    \label{eq:hamiltonian_base}
\end{equation}

\subsection{Superselection sectors}%
\label{sub:topological_sectors}

\begin{figure}[t]
    \centering
    \includegraphics{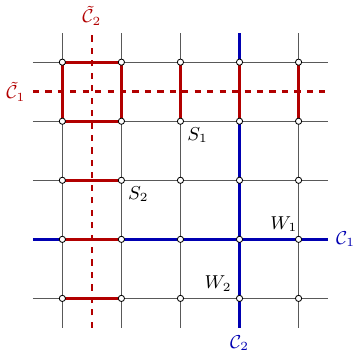}
    \caption{Graphical representation of the non-local order parameters $\overline{W}_{1,2}$ (in blue) and $\overline{S}_{1,2}$ (in red) and their respective paths $\mathcal{C}_{1,2}$ and $\tilde{\mathcal{C}}_{1,2}$.}
    \label{fig:nonlocal_operators}
\end{figure}

One of the main features of \eqref{eq:hamiltonian_base} is the presence of topologically protected superselection sectors.
In order to illustrate this, we first need to define the non-contractible Wilson loop operators (pictured in blue in Fig.~\ref{fig:nonlocal_operators}):
\begin{equation}
    \overline{W}_i = \prod_{\ell \in \mathcal{C}_i} U_\ell, \qquad i=1,2,
    \label{eq:top_wilson_loop}
\end{equation}
where $\mathcal{C}_i$ are non-contractible loops around the lattice $\mathbb{L}$, along the $\hat{i}$ direction.
A simple calculation shows that both $\overline{W}_1$ and $\overline{W}_2$ commute with all $G_x$, thus they are gauge-invariant, but one also finds out that none of them can be written as a product of $U_{\square}$ nor $V_\ell$.

Besides $\overline{W}_i$, another type of non-local operators have to be introduced.
They are defined on \emph{cuts} of the lattice $\L$, i.e.~paths on the dual lattice $\tilde{\L}$.
Consider \emph{non-contractible} cuts $\tilde{\mathcal{C}}_1$ and $\tilde{\mathcal{C}}_2$ along the directions $\hat{1}$ and $\hat{2}$, respectively.
On this cuts, the 't Hooft string operators $\overline{S}_1$ and $\overline{S}_2$ are constructed as
\begin{equation}
    \overline{S}_i = \prod_{\ell \in \tilde{\mathcal{C}}_i} V_\ell, \qquad i=1,2,
    \label{eq:top_string_operators}
\end{equation}
in a similar fashion to \eqref{eq:top_wilson_loop} (shown in red in Fig.~\ref{fig:nonlocal_operators}).



These two classes of non-local operators resembles the same operators of the Toric Code \cite{kitaev2003fault}, that distinguish the degenerate ground states.
One key difference here is that the operators $\overline{W}_i$ do not commute with the Hamiltonian \eqref{eq:hamiltonian_base}, which contains an electric field term.
Thus, unlike the Toric Code, we no longer have degenerate ground states when $\lambda \neq 0$.
But we can still use the $\overline{S}_i$ operators to decompose the Hilbert space $\Hphys$, since they still commute with all the \emph{local operators} $U_{\square}$ and $V_{\ell}$ (thus also with $H_{\Z_N}$).
Now one can see that the operator $\overline{S}_i$ ($i=1,2$) of \eqref{eq:top_string_operators} has $N$ eigenvalues $\omega^n$, with $n=1, \dots, N-1$.
Hence, one can decompose $\Hphys$ as sum of superselection sectors
\begin{equation}
    \Hphys = \bigoplus_{n, m=0}^{N-1} \Hphys^{(n, m)},
    \label{eq:decomposizione_Hphys}
\end{equation}
where for each $\ket{\phi} \in \Hphys^{(n, m)}$ we have:
\begin{equation}
    \overline{S}_1 \ket{\phi} = \omega^{m}\ket{\phi}, \quad
    \overline{S}_2 \ket{\phi} = \omega^{n}\ket{\phi}.
\end{equation}
Let us consider now the role of the Wilson loops $\overline{W}_i$.
One can see that:
\begin{equation}
    \overline{W}_2 \overline{S}_1 = \omega \overline{S}_1 \overline{W}_2, \qquad
    \overline{W}_1 \overline{S}_2 = \omega \overline{S}_2 \overline{W}_1.
    \label{eq:algebra_op_nonlocali}
\end{equation}
It follows that $\overline{W}_{1,2}$ acts as a shift operator for the eigenspaces of $\overline{S}_{2,1}$:
\begin{equation}
    \overline{W}_1 : \Hphys^{(n, m)} \to \Hphys^{(n + 1, m)}, \quad
    \overline{W}_2 : \Hphys^{(n, m)} \to \Hphys^{(n, m + 1)},
    \label{eq:azione_wilson_loop}
\end{equation}
where the integers $n + 1$ and $m + 1$ have to be taken $\mathrm{mod}\; N$.

From a physical point of view, the Wilson loops operators $\overline{W}_1$ and $\overline{W}_2$ create non-contractible electric loops around the lattice, while the 't Hooft strings $\overline{S}_2$ and $\overline{S}_1$ detect the presence and the strength of these electric loops.
Therefore, it is clear that the Hilbert subspace $\Hphys^{(n, m)}$ is the subspace of all the states that contains an electric loop of strength $\omega^n$ and $\omega^{m}$ along the $\hat{1}$ and $\hat{2}$ direction, respectively.
Furthermore, the evolution of a state in $\Hphys^{(n,m)}$ with the Hamiltonian in \eqref{eq:hamiltonian_base} is confined in $\Hphys^{(n,m)}$.

\begin{figure*}[t]
	\centering
	\hspace{3em}$\Z_2$ g.s.~amplitudes distribution, $\lambda=0.1$ \\[5pt]
	\includegraphics{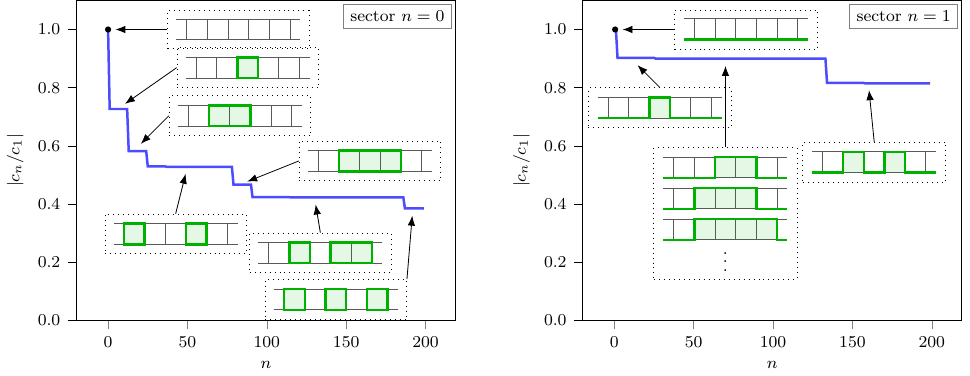}
	\caption{
		$\Z_2$ ground state amplitude distribution for $\lambda=0.1$ of the first 200 states and with lattice size $12 \times 2$.
		\emph{Top}: distribution of the ratios $|{c_n/c_1}|$ for the sector $n=0$ (see \eqref{eq:gs_amplitudes}).
		We see that the heaviest states that enter the ground state, apart from the vacuum that sets the scale, are made of small electric loops, typical of a confined phase.
		\emph{Bottom}: the same distribution of ratios for the sector $n=1$.
		We see that the heaviest states are made of bigger and bigger deformations of the electric string that goes around the ladder.
		This happens because the energy contributions depend only on the number of domain walls between two regions with different flux content.
                The length of each step in the (blue) curves can be found by means of combinatorial methods.
                For example, the first step in the left panel corresponds to the 12 possible states contain just one closed loop encircling a single plaquette.
	}
	\label{fig:gs_ampl_distr_0.1_Z2}
\end{figure*}

\begin{figure*}[t]
	\centering
	\hspace{3em}$\Z_2$ g.s.~amplitudes distribution, $\lambda=1.5$\\[5pt]
	\includegraphics{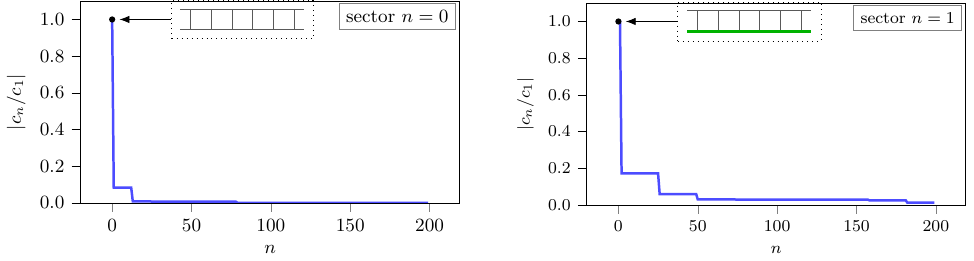}
	\caption{
		$\Z_2$ ground state amplitude distribution for $\lambda=1.5$ of the first 200 states and with lattice size $12 \times 2$.
		For both sectors $n=0$ (\emph{top}) and $n=1$ (\emph{bottom}) we are in a confined phase, which corresponds to a ferromagnetic phase in the Ising chain.
		Here we see a polarized state where the domain walls are suppressed and the ground state is essentially a product state.
	}
	\label{fig:gs_ampl_distr_1.5_Z2}
\end{figure*}

\section{Distribution of the amplitudes in the ground state}
\label{app:ground_state_distribution}

In the $N=2$ case, we further differentiate the phase diagrams of the two sectors by looking at the ground state amplitudes distribution, for $\lambda<1$ and $\lambda>1$.
The ground state can be written as a superposition of the gauge invariant states of $\Hphys$ in the given sector
\begin{equation}
    \ket{\Psi_{\text{g.s.} }}= \sum_n c_n \ket{n}.
    \label{eq:gs_amplitudes}
\end{equation}
The basis $\ket{n}$ and the amplitudes $c_n$ are sorted in a decreasing order with respect to their modulus.
The first state of the list, with amplitude $c_1$, is always the Fock vacua $\ket{\Omega_n}$ of the sector $n$, hence we consider the distribution of the ratios $\abs{c_n / c_1}$, which are plotted in Fig.~\ref{fig:gs_ampl_distr_0.1_Z2}--\ref{fig:gs_ampl_distr_1.5_Z2} for $\lambda=0.1$ and $\lambda=1.5$, respectively.

The most interesting one is at $\lambda = 0.1$ in Fig.~\ref{fig:gs_ampl_distr_0.1_Z2}, where the difference between the deconfined phase in the sector $n=1$ and the confined one in the sector $n=0$ can be seen.
In particular, in the sector $n=1$ the ground state is a superposition of deformations of the non-contractible electric loop (that makes the Fock vacuum).
For this reason, the ground state can be thought as a \emph{kink condensate} \cite{fradkin1978order} (which is a paramagnetic state), where each kink corresponds to a deformation of the electric loop.
This behaviour differs from the confined state of the sector $n=0$, where most of the contributions to the ground state comes from states with small electric loops.
In other words, the creation of magnetic fluxes is suppressed.
The peculiar behaviour of the sector $n=1$ is due to the fact that the energy contributions depend only on the domain walls between two plaquettes with different flux content.
While in the sector $n=0$ we also receive energy contributions from the presence of the fluxes themselves.

Meanwhile, for $\lambda > 1$, where we have confinement in both sectors, the ground state is essentially a product state, akin to a ferromagnetic state, like is shown in Fig.~\ref{fig:gs_ampl_distr_1.5_Z2}.

\bibliography{biblio.bib}

\begin{thebibliography}{49}%
\makeatletter
\providecommand \@ifxundefined [1]{%
 \@ifx{#1\undefined}
}%
\providecommand \@ifnum [1]{%
 \ifnum #1\expandafter \@firstoftwo
 \else \expandafter \@secondoftwo
 \fi
}%
\providecommand \@ifx [1]{%
 \ifx #1\expandafter \@firstoftwo
 \else \expandafter \@secondoftwo
 \fi
}%
\providecommand \natexlab [1]{#1}%
\providecommand \enquote  [1]{``#1''}%
\providecommand \bibnamefont  [1]{#1}%
\providecommand \bibfnamefont [1]{#1}%
\providecommand \citenamefont [1]{#1}%
\providecommand \href@noop [0]{\@secondoftwo}%
\providecommand \href [0]{\begingroup \@sanitize@url \@href}%
\providecommand \@href[1]{\@@startlink{#1}\@@href}%
\providecommand \@@href[1]{\endgroup#1\@@endlink}%
\providecommand \@sanitize@url [0]{\catcode `\\12\catcode `\$12\catcode `\&12\catcode `\#12\catcode `\^12\catcode `\_12\catcode `\%12\relax}%
\providecommand \@@startlink[1]{}%
\providecommand \@@endlink[0]{}%
\providecommand \url  [0]{\begingroup\@sanitize@url \@url }%
\providecommand \@url [1]{\endgroup\@href {#1}{\urlprefix }}%
\providecommand \urlprefix  [0]{URL }%
\providecommand \Eprint [0]{\href }%
\providecommand \doibase [0]{http://dx.doi.org/}%
\providecommand \selectlanguage [0]{\@gobble}%
\providecommand \bibinfo  [0]{\@secondoftwo}%
\providecommand \bibfield  [0]{\@secondoftwo}%
\providecommand \translation [1]{[#1]}%
\providecommand \BibitemOpen [0]{}%
\providecommand \bibitemStop [0]{}%
\providecommand \bibitemNoStop [0]{.\EOS\space}%
\providecommand \EOS [0]{\spacefactor3000\relax}%
\providecommand \BibitemShut  [1]{\csname bibitem#1\endcsname}%
\let\auto@bib@innerbib\@empty
\bibitem [{\citenamefont {Wilson}(1974)}]{wilson1974confinement}%
  \BibitemOpen
  \bibfield  {author} {\bibinfo {author} {\bibfnamefont {K.~G.}\ \bibnamefont {Wilson}},\ }\href {\doibase 10.1103/PhysRevD.10.2445} {\bibfield  {journal} {\bibinfo  {journal} {Phys. Rev. D}\ }\textbf {\bibinfo {volume} {10}},\ \bibinfo {pages} {2445} (\bibinfo {year} {1974})}\BibitemShut {NoStop}%
\bibitem [{\citenamefont {Kogut}\ and\ \citenamefont {Susskind}(1975)}]{kogut1975hamiltonian}%
  \BibitemOpen
  \bibfield  {author} {\bibinfo {author} {\bibfnamefont {J.}~\bibnamefont {Kogut}}\ and\ \bibinfo {author} {\bibfnamefont {L.}~\bibnamefont {Susskind}},\ }\href {\doibase 10.1103/PhysRevD.11.395} {\bibfield  {journal} {\bibinfo  {journal} {Phys. Rev. D}\ }\textbf {\bibinfo {volume} {11}},\ \bibinfo {pages} {395} (\bibinfo {year} {1975})}\BibitemShut {NoStop}%
\bibitem [{\citenamefont {Kogut}(1979)}]{kogut1979introduction}%
  \BibitemOpen
  \bibfield  {author} {\bibinfo {author} {\bibfnamefont {J.~B.}\ \bibnamefont {Kogut}},\ }\href {\doibase 10.1103/RevModPhys.51.659} {\bibfield  {journal} {\bibinfo  {journal} {Rev. Mod. Phys.}\ }\textbf {\bibinfo {volume} {51}},\ \bibinfo {pages} {659} (\bibinfo {year} {1979})}\BibitemShut {NoStop}%
\bibitem [{\citenamefont {Feynman}(2018)}]{feynman2018simulating}%
  \BibitemOpen
  \bibfield  {author} {\bibinfo {author} {\bibfnamefont {R.~P.}\ \bibnamefont {Feynman}},\ }in\ \href@noop {} {\emph {\bibinfo {booktitle} {Feynman and computation}}}\ (\bibinfo  {publisher} {CRC Press},\ \bibinfo {year} {2018})\ pp.\ \bibinfo {pages} {133--153}\BibitemShut {NoStop}%
\bibitem [{\citenamefont {Feynman}(1985)}]{feynman1985quantum}%
  \BibitemOpen
  \bibfield  {author} {\bibinfo {author} {\bibfnamefont {R.~P.}\ \bibnamefont {Feynman}},\ }\href {\doibase 10.1364/ON.11.2.000011} {\bibfield  {journal} {\bibinfo  {journal} {Optics News}\ }\textbf {\bibinfo {volume} {11}},\ \bibinfo {pages} {11} (\bibinfo {year} {1985})}\BibitemShut {NoStop}%
\bibitem [{\citenamefont {Ba{\~n}uls}\ \emph {et~al.}(2020)\citenamefont {Ba{\~n}uls}, \citenamefont {Blatt}, \citenamefont {Catani}, \citenamefont {Celi}, \citenamefont {Cirac}, \citenamefont {Dalmonte}, \citenamefont {Fallani}, \citenamefont {Jansen}, \citenamefont {Lewenstein}, \citenamefont {Montangero} \emph {et~al.}}]{banuls2020simulating}%
  \BibitemOpen
  \bibfield  {author} {\bibinfo {author} {\bibfnamefont {M.~C.}\ \bibnamefont {Ba{\~n}uls}}, \bibinfo {author} {\bibfnamefont {R.}~\bibnamefont {Blatt}}, \bibinfo {author} {\bibfnamefont {J.}~\bibnamefont {Catani}}, \bibinfo {author} {\bibfnamefont {A.}~\bibnamefont {Celi}}, \bibinfo {author} {\bibfnamefont {J.~I.}\ \bibnamefont {Cirac}}, \bibinfo {author} {\bibfnamefont {M.}~\bibnamefont {Dalmonte}}, \bibinfo {author} {\bibfnamefont {L.}~\bibnamefont {Fallani}}, \bibinfo {author} {\bibfnamefont {K.}~\bibnamefont {Jansen}}, \bibinfo {author} {\bibfnamefont {M.}~\bibnamefont {Lewenstein}}, \bibinfo {author} {\bibfnamefont {S.}~\bibnamefont {Montangero}},  \emph {et~al.},\ }\href {\doibase 10.1140/epjd/e2020-100571-8} {\bibfield  {journal} {\bibinfo  {journal} {The European Physical Journal D}\ }\textbf {\bibinfo {volume} {74}},\ \bibinfo {pages} {1} (\bibinfo {year} {2020})}\BibitemShut {NoStop}%
\bibitem [{\citenamefont {Zohar}\ \emph {et~al.}(2015)\citenamefont {Zohar}, \citenamefont {Cirac},\ and\ \citenamefont {Reznik}}]{zohar2015quantum}%
  \BibitemOpen
  \bibfield  {author} {\bibinfo {author} {\bibfnamefont {E.}~\bibnamefont {Zohar}}, \bibinfo {author} {\bibfnamefont {J.~I.}\ \bibnamefont {Cirac}}, \ and\ \bibinfo {author} {\bibfnamefont {B.}~\bibnamefont {Reznik}},\ }\href {\doibase 10.1088/0034-4885/79/1/014401} {\bibfield  {journal} {\bibinfo  {journal} {Reports on Progress in Physics}\ }\textbf {\bibinfo {volume} {79}},\ \bibinfo {pages} {014401} (\bibinfo {year} {2015})}\BibitemShut {NoStop}%
\bibitem [{\citenamefont {Dalmonte}\ and\ \citenamefont {Montangero}(2016)}]{dalmonte2016latticegauge}%
  \BibitemOpen
  \bibfield  {author} {\bibinfo {author} {\bibfnamefont {M.}~\bibnamefont {Dalmonte}}\ and\ \bibinfo {author} {\bibfnamefont {S.}~\bibnamefont {Montangero}},\ }\href {\doibase 10.1080/00107514.2016.1151199} {\bibfield  {journal} {\bibinfo  {journal} {Contemporary Physics}\ }\textbf {\bibinfo {volume} {57}},\ \bibinfo {pages} {388} (\bibinfo {year} {2016})}\BibitemShut {NoStop}%
\bibitem [{\citenamefont {Zohar}\ and\ \citenamefont {Burrello}(2015)}]{zohar2015latticegauge}%
  \BibitemOpen
  \bibfield  {author} {\bibinfo {author} {\bibfnamefont {E.}~\bibnamefont {Zohar}}\ and\ \bibinfo {author} {\bibfnamefont {M.}~\bibnamefont {Burrello}},\ }\href {\doibase 10.1103/PhysRevD.91.054506} {\bibfield  {journal} {\bibinfo  {journal} {Phys. Rev. D}\ }\textbf {\bibinfo {volume} {91}},\ \bibinfo {pages} {054506} (\bibinfo {year} {2015})}\BibitemShut {NoStop}%
\bibitem [{\citenamefont {Zohar}\ \emph {et~al.}(2017{\natexlab{a}})\citenamefont {Zohar}, \citenamefont {Farace}, \citenamefont {Reznik},\ and\ \citenamefont {Cirac}}]{zohar2017digital}%
  \BibitemOpen
  \bibfield  {author} {\bibinfo {author} {\bibfnamefont {E.}~\bibnamefont {Zohar}}, \bibinfo {author} {\bibfnamefont {A.}~\bibnamefont {Farace}}, \bibinfo {author} {\bibfnamefont {B.}~\bibnamefont {Reznik}}, \ and\ \bibinfo {author} {\bibfnamefont {J.~I.}\ \bibnamefont {Cirac}},\ }\href {\doibase 10.1103/PhysRevA.95.023604} {\bibfield  {journal} {\bibinfo  {journal} {Phys. Rev. A}\ }\textbf {\bibinfo {volume} {95}},\ \bibinfo {pages} {023604} (\bibinfo {year} {2017}{\natexlab{a}})}\BibitemShut {NoStop}%
\bibitem [{\citenamefont {Fradkin}\ and\ \citenamefont {Shenker}(1979)}]{fradkin1979phase}%
  \BibitemOpen
  \bibfield  {author} {\bibinfo {author} {\bibfnamefont {E.}~\bibnamefont {Fradkin}}\ and\ \bibinfo {author} {\bibfnamefont {S.~H.}\ \bibnamefont {Shenker}},\ }\href {\doibase 10.1103/PhysRevD.19.3682} {\bibfield  {journal} {\bibinfo  {journal} {Phys. Rev. D}\ }\textbf {\bibinfo {volume} {19}},\ \bibinfo {pages} {3682} (\bibinfo {year} {1979})}\BibitemShut {NoStop}%
\bibitem [{\citenamefont {Horn}\ \emph {et~al.}(1979)\citenamefont {Horn}, \citenamefont {Weinstein},\ and\ \citenamefont {Yankielowicz}}]{horn1979zngauge}%
  \BibitemOpen
  \bibfield  {author} {\bibinfo {author} {\bibfnamefont {D.}~\bibnamefont {Horn}}, \bibinfo {author} {\bibfnamefont {M.}~\bibnamefont {Weinstein}}, \ and\ \bibinfo {author} {\bibfnamefont {S.}~\bibnamefont {Yankielowicz}},\ }\href {\doibase 10.1103/PhysRevD.19.3715} {\bibfield  {journal} {\bibinfo  {journal} {Phys. Rev. D}\ }\textbf {\bibinfo {volume} {19}},\ \bibinfo {pages} {3715} (\bibinfo {year} {1979})}\BibitemShut {NoStop}%
\bibitem [{\citenamefont {Tagliacozzo}\ and\ \citenamefont {Vidal}(2011)}]{tagliacozzo2011entanglement}%
  \BibitemOpen
  \bibfield  {author} {\bibinfo {author} {\bibfnamefont {L.}~\bibnamefont {Tagliacozzo}}\ and\ \bibinfo {author} {\bibfnamefont {G.}~\bibnamefont {Vidal}},\ }\href {\doibase 10.1103/PhysRevB.83.115127} {\bibfield  {journal} {\bibinfo  {journal} {Phys. Rev. B}\ }\textbf {\bibinfo {volume} {83}},\ \bibinfo {pages} {115127} (\bibinfo {year} {2011})}\BibitemShut {NoStop}%
\bibitem [{\citenamefont {Tagliacozzo}\ \emph {et~al.}(2013)\citenamefont {Tagliacozzo}, \citenamefont {Celi}, \citenamefont {Zamora},\ and\ \citenamefont {Lewenstein}}]{tagliacozzo2013optical}%
  \BibitemOpen
  \bibfield  {author} {\bibinfo {author} {\bibfnamefont {L.}~\bibnamefont {Tagliacozzo}}, \bibinfo {author} {\bibfnamefont {A.}~\bibnamefont {Celi}}, \bibinfo {author} {\bibfnamefont {A.}~\bibnamefont {Zamora}}, \ and\ \bibinfo {author} {\bibfnamefont {M.}~\bibnamefont {Lewenstein}},\ }\href {\doibase https://doi.org/10.1016/j.aop.2012.11.009} {\bibfield  {journal} {\bibinfo  {journal} {Annals of Physics}\ }\textbf {\bibinfo {volume} {330}},\ \bibinfo {pages} {160} (\bibinfo {year} {2013})}\BibitemShut {NoStop}%
\bibitem [{\citenamefont {Hamma}\ and\ \citenamefont {Lidar}(2008)}]{hamma2008adiabatic}%
  \BibitemOpen
  \bibfield  {author} {\bibinfo {author} {\bibfnamefont {A.}~\bibnamefont {Hamma}}\ and\ \bibinfo {author} {\bibfnamefont {D.~A.}\ \bibnamefont {Lidar}},\ }\href {\doibase 10.1103/PhysRevLett.100.030502} {\bibfield  {journal} {\bibinfo  {journal} {Phys. Rev. Lett.}\ }\textbf {\bibinfo {volume} {100}},\ \bibinfo {pages} {030502} (\bibinfo {year} {2008})}\BibitemShut {NoStop}%
\bibitem [{\citenamefont {Trebst}\ \emph {et~al.}(2007)\citenamefont {Trebst}, \citenamefont {Werner}, \citenamefont {Troyer}, \citenamefont {Shtengel},\ and\ \citenamefont {Nayak}}]{trebst2007topological}%
  \BibitemOpen
  \bibfield  {author} {\bibinfo {author} {\bibfnamefont {S.}~\bibnamefont {Trebst}}, \bibinfo {author} {\bibfnamefont {P.}~\bibnamefont {Werner}}, \bibinfo {author} {\bibfnamefont {M.}~\bibnamefont {Troyer}}, \bibinfo {author} {\bibfnamefont {K.}~\bibnamefont {Shtengel}}, \ and\ \bibinfo {author} {\bibfnamefont {C.}~\bibnamefont {Nayak}},\ }\href {\doibase 10.1103/PhysRevLett.98.070602} {\bibfield  {journal} {\bibinfo  {journal} {Phys. Rev. Lett.}\ }\textbf {\bibinfo {volume} {98}},\ \bibinfo {pages} {070602} (\bibinfo {year} {2007})}\BibitemShut {NoStop}%
\bibitem [{\citenamefont {Emonts}\ \emph {et~al.}(2020)\citenamefont {Emonts}, \citenamefont {Ba\~nuls}, \citenamefont {Cirac},\ and\ \citenamefont {Zohar}}]{emonts2020z3gauge}%
  \BibitemOpen
  \bibfield  {author} {\bibinfo {author} {\bibfnamefont {P.}~\bibnamefont {Emonts}}, \bibinfo {author} {\bibfnamefont {M.~C.}\ \bibnamefont {Ba\~nuls}}, \bibinfo {author} {\bibfnamefont {I.}~\bibnamefont {Cirac}}, \ and\ \bibinfo {author} {\bibfnamefont {E.}~\bibnamefont {Zohar}},\ }\href {\doibase 10.1103/PhysRevD.102.074501} {\bibfield  {journal} {\bibinfo  {journal} {Phys. Rev. D}\ }\textbf {\bibinfo {volume} {102}},\ \bibinfo {pages} {074501} (\bibinfo {year} {2020})}\BibitemShut {NoStop}%
\bibitem [{\citenamefont {Zohar}\ \emph {et~al.}(2017{\natexlab{b}})\citenamefont {Zohar}, \citenamefont {Farace}, \citenamefont {Reznik},\ and\ \citenamefont {Cirac}}]{zohar2017z2gauge}%
  \BibitemOpen
  \bibfield  {author} {\bibinfo {author} {\bibfnamefont {E.}~\bibnamefont {Zohar}}, \bibinfo {author} {\bibfnamefont {A.}~\bibnamefont {Farace}}, \bibinfo {author} {\bibfnamefont {B.}~\bibnamefont {Reznik}}, \ and\ \bibinfo {author} {\bibfnamefont {J.~I.}\ \bibnamefont {Cirac}},\ }\href {\doibase 10.1103/PhysRevLett.118.070501} {\bibfield  {journal} {\bibinfo  {journal} {Phys. Rev. Lett.}\ }\textbf {\bibinfo {volume} {118}},\ \bibinfo {pages} {070501} (\bibinfo {year} {2017}{\natexlab{b}})}\BibitemShut {NoStop}%
\bibitem [{\citenamefont {Nyhegn}\ \emph {et~al.}(2021)\citenamefont {Nyhegn}, \citenamefont {Chung},\ and\ \citenamefont {Burrello}}]{burrello2021ladder}%
  \BibitemOpen
  \bibfield  {author} {\bibinfo {author} {\bibfnamefont {J.}~\bibnamefont {Nyhegn}}, \bibinfo {author} {\bibfnamefont {C.-M.}\ \bibnamefont {Chung}}, \ and\ \bibinfo {author} {\bibfnamefont {M.}~\bibnamefont {Burrello}},\ }\href {\doibase 10.1103/PhysRevResearch.3.013133} {\bibfield  {journal} {\bibinfo  {journal} {Phys. Rev. Research}\ }\textbf {\bibinfo {volume} {3}},\ \bibinfo {pages} {013133} (\bibinfo {year} {2021})}\BibitemShut {NoStop}%
\bibitem [{\citenamefont {Cobanera}\ \emph {et~al.}(2011)\citenamefont {Cobanera}, \citenamefont {Ortiz},\ and\ \citenamefont {Nussinov}}]{cobanera2011bond}%
  \BibitemOpen
  \bibfield  {author} {\bibinfo {author} {\bibfnamefont {E.}~\bibnamefont {Cobanera}}, \bibinfo {author} {\bibfnamefont {G.}~\bibnamefont {Ortiz}}, \ and\ \bibinfo {author} {\bibfnamefont {Z.}~\bibnamefont {Nussinov}},\ }\href {\doibase 10.1080/00018732.2011.619814} {\bibfield  {journal} {\bibinfo  {journal} {Advances in physics}\ }\textbf {\bibinfo {volume} {60}},\ \bibinfo {pages} {679} (\bibinfo {year} {2011})}\BibitemShut {NoStop}%
\bibitem [{\citenamefont {Nussinov}\ and\ \citenamefont {Ortiz}(2009)}]{nussinov2009bond}%
  \BibitemOpen
  \bibfield  {author} {\bibinfo {author} {\bibfnamefont {Z.}~\bibnamefont {Nussinov}}\ and\ \bibinfo {author} {\bibfnamefont {G.}~\bibnamefont {Ortiz}},\ }\href {https://link.aps.org/doi/10.1103/PhysRevB.79.214440} {\bibfield  {journal} {\bibinfo  {journal} {Phys. Rev. B}\ }\textbf {\bibinfo {volume} {79}},\ \bibinfo {pages} {214440} (\bibinfo {year} {2009})}\BibitemShut {NoStop}%
\bibitem [{\citenamefont {Baxter}(1989)}]{baxter1989clock}%
  \BibitemOpen
  \bibfield  {author} {\bibinfo {author} {\bibfnamefont {R.~J.}\ \bibnamefont {Baxter}},\ }\href {\doibase https://doi.org/10.1016/0375-9601(89)90884-0} {\bibfield  {journal} {\bibinfo  {journal} {Physics Letters A}\ }\textbf {\bibinfo {volume} {140}},\ \bibinfo {pages} {155} (\bibinfo {year} {1989})}\BibitemShut {NoStop}%
\bibitem [{\citenamefont {Ortiz}\ \emph {et~al.}(2012)\citenamefont {Ortiz}, \citenamefont {Cobanera},\ and\ \citenamefont {Nussinov}}]{ortiz2012dualities}%
  \BibitemOpen
  \bibfield  {author} {\bibinfo {author} {\bibfnamefont {G.}~\bibnamefont {Ortiz}}, \bibinfo {author} {\bibfnamefont {E.}~\bibnamefont {Cobanera}}, \ and\ \bibinfo {author} {\bibfnamefont {Z.}~\bibnamefont {Nussinov}},\ }\href {\doibase https://doi.org/10.1016/j.nuclphysb.2011.09.012} {\bibfield  {journal} {\bibinfo  {journal} {Nuclear Physics B}\ }\textbf {\bibinfo {volume} {854}},\ \bibinfo {pages} {780} (\bibinfo {year} {2012})}\BibitemShut {NoStop}%
\bibitem [{\citenamefont {Fendley}(2014)}]{fendley2014parafermions}%
  \BibitemOpen
  \bibfield  {author} {\bibinfo {author} {\bibfnamefont {P.}~\bibnamefont {Fendley}},\ }\href {\doibase 10.1088/1751-8113/47/7/075001} {\bibfield  {journal} {\bibinfo  {journal} {Journal of Physics A: Mathematical and Theoretical}\ }\textbf {\bibinfo {volume} {47}},\ \bibinfo {pages} {075001} (\bibinfo {year} {2014})}\BibitemShut {NoStop}%
\bibitem [{\citenamefont {Zhuang}\ \emph {et~al.}(2015)\citenamefont {Zhuang}, \citenamefont {Changlani}, \citenamefont {Tubman},\ and\ \citenamefont {Hughes}}]{zhuang2015clock}%
  \BibitemOpen
  \bibfield  {author} {\bibinfo {author} {\bibfnamefont {Y.}~\bibnamefont {Zhuang}}, \bibinfo {author} {\bibfnamefont {H.~J.}\ \bibnamefont {Changlani}}, \bibinfo {author} {\bibfnamefont {N.~M.}\ \bibnamefont {Tubman}}, \ and\ \bibinfo {author} {\bibfnamefont {T.~L.}\ \bibnamefont {Hughes}},\ }\href {\doibase 10.1103/PhysRevB.92.035154} {\bibfield  {journal} {\bibinfo  {journal} {Phys. Rev. B}\ }\textbf {\bibinfo {volume} {92}},\ \bibinfo {pages} {035154} (\bibinfo {year} {2015})}\BibitemShut {NoStop}%
\bibitem [{\citenamefont {Sun}\ \emph {et~al.}(2019)\citenamefont {Sun}, \citenamefont {Vekua}, \citenamefont {Cobanera},\ and\ \citenamefont {Ortiz}}]{sun2019phase}%
  \BibitemOpen
  \bibfield  {author} {\bibinfo {author} {\bibfnamefont {G.}~\bibnamefont {Sun}}, \bibinfo {author} {\bibfnamefont {T.}~\bibnamefont {Vekua}}, \bibinfo {author} {\bibfnamefont {E.}~\bibnamefont {Cobanera}}, \ and\ \bibinfo {author} {\bibfnamefont {G.}~\bibnamefont {Ortiz}},\ }\href {\doibase 10.1103/PhysRevB.100.094428} {\bibfield  {journal} {\bibinfo  {journal} {Phys. Rev. B}\ }\textbf {\bibinfo {volume} {100}},\ \bibinfo {pages} {094428} (\bibinfo {year} {2019})}\BibitemShut {NoStop}%
\bibitem [{\citenamefont {Schwinger}(1960)}]{schwinger1960unitary}%
  \BibitemOpen
  \bibfield  {author} {\bibinfo {author} {\bibfnamefont {J.}~\bibnamefont {Schwinger}},\ }\href {\doibase 10.1073/pnas.46.4.570} {\bibfield  {journal} {\bibinfo  {journal} {Proceedings of the National Academy of Sciences}\ }\textbf {\bibinfo {volume} {46}},\ \bibinfo {pages} {570} (\bibinfo {year} {1960})}\BibitemShut {NoStop}%
\bibitem [{\citenamefont {Notarnicola}\ \emph {et~al.}(2015)\citenamefont {Notarnicola}, \citenamefont {Ercolessi}, \citenamefont {Facchi}, \citenamefont {Marmo}, \citenamefont {Pascazio},\ and\ \citenamefont {Pepe}}]{notarnicola2015discrete}%
  \BibitemOpen
  \bibfield  {author} {\bibinfo {author} {\bibfnamefont {S.}~\bibnamefont {Notarnicola}}, \bibinfo {author} {\bibfnamefont {E.}~\bibnamefont {Ercolessi}}, \bibinfo {author} {\bibfnamefont {P.}~\bibnamefont {Facchi}}, \bibinfo {author} {\bibfnamefont {G.}~\bibnamefont {Marmo}}, \bibinfo {author} {\bibfnamefont {S.}~\bibnamefont {Pascazio}}, \ and\ \bibinfo {author} {\bibfnamefont {F.~V.}\ \bibnamefont {Pepe}},\ }\href {\doibase 10.1088/1751-8113/48/30/30ft01} {\bibfield  {journal} {\bibinfo  {journal} {Journal of Physics A: Mathematical and Theoretical}\ }\textbf {\bibinfo {volume} {48}},\ \bibinfo {pages} {30FT01} (\bibinfo {year} {2015})}\BibitemShut {NoStop}%
\bibitem [{\citenamefont {Ercolessi}\ \emph {et~al.}(2018)\citenamefont {Ercolessi}, \citenamefont {Facchi}, \citenamefont {Magnifico}, \citenamefont {Pascazio},\ and\ \citenamefont {Pepe}}]{ercolessi2018znmodels}%
  \BibitemOpen
  \bibfield  {author} {\bibinfo {author} {\bibfnamefont {E.}~\bibnamefont {Ercolessi}}, \bibinfo {author} {\bibfnamefont {P.}~\bibnamefont {Facchi}}, \bibinfo {author} {\bibfnamefont {G.}~\bibnamefont {Magnifico}}, \bibinfo {author} {\bibfnamefont {S.}~\bibnamefont {Pascazio}}, \ and\ \bibinfo {author} {\bibfnamefont {F.~V.}\ \bibnamefont {Pepe}},\ }\href {\doibase 10.1103/PhysRevD.98.074503} {\bibfield  {journal} {\bibinfo  {journal} {Phys. Rev. D}\ }\textbf {\bibinfo {volume} {98}},\ \bibinfo {pages} {074503} (\bibinfo {year} {2018})}\BibitemShut {NoStop}%
\bibitem [{\citenamefont {Magnifico}\ \emph {et~al.}(2020)\citenamefont {Magnifico}, \citenamefont {Dalmonte}, \citenamefont {Facchi}, \citenamefont {Pascazio}, \citenamefont {Pepe},\ and\ \citenamefont {Ercolessi}}]{magnifico2020realtimedynamics}%
  \BibitemOpen
  \bibfield  {author} {\bibinfo {author} {\bibfnamefont {G.}~\bibnamefont {Magnifico}}, \bibinfo {author} {\bibfnamefont {M.}~\bibnamefont {Dalmonte}}, \bibinfo {author} {\bibfnamefont {P.}~\bibnamefont {Facchi}}, \bibinfo {author} {\bibfnamefont {S.}~\bibnamefont {Pascazio}}, \bibinfo {author} {\bibfnamefont {F.~V.}\ \bibnamefont {Pepe}}, \ and\ \bibinfo {author} {\bibfnamefont {E.}~\bibnamefont {Ercolessi}},\ }\href {\doibase 10.22331/q-2020-06-15-281} {\bibfield  {journal} {\bibinfo  {journal} {{Quantum}}\ }\textbf {\bibinfo {volume} {4}},\ \bibinfo {pages} {281} (\bibinfo {year} {2020})}\BibitemShut {NoStop}%
\bibitem [{\citenamefont {Weyl}(1950)}]{weyl1950theory}%
  \BibitemOpen
  \bibfield  {author} {\bibinfo {author} {\bibfnamefont {H.}~\bibnamefont {Weyl}},\ }\href@noop {} {\emph {\bibinfo {title} {The theory of groups and quantum mechanics}}}\ (\bibinfo  {publisher} {Courier Corporation},\ \bibinfo {year} {1950})\BibitemShut {NoStop}%
\bibitem [{\citenamefont {Kitaev}(2003)}]{kitaev2003fault}%
  \BibitemOpen
  \bibfield  {author} {\bibinfo {author} {\bibfnamefont {A.~Y.}\ \bibnamefont {Kitaev}},\ }\href {\doibase https://doi.org/10.1016/S0003-4916(02)00018-0} {\bibfield  {journal} {\bibinfo  {journal} {Annals of Physics}\ }\textbf {\bibinfo {volume} {303}},\ \bibinfo {pages} {2} (\bibinfo {year} {2003})}\BibitemShut {NoStop}%
\bibitem [{\citenamefont {Fradkin}\ and\ \citenamefont {Susskind}(1978)}]{fradkin1978order}%
  \BibitemOpen
  \bibfield  {author} {\bibinfo {author} {\bibfnamefont {E.}~\bibnamefont {Fradkin}}\ and\ \bibinfo {author} {\bibfnamefont {L.}~\bibnamefont {Susskind}},\ }\href {\doibase 10.1103/PhysRevD.17.2637} {\bibfield  {journal} {\bibinfo  {journal} {Phys. Rev. D}\ }\textbf {\bibinfo {volume} {17}},\ \bibinfo {pages} {2637} (\bibinfo {year} {1978})}\BibitemShut {NoStop}%
\bibitem [{Note1()}]{Note1}%
  \BibitemOpen
  \bibinfo {note} {Thanks to \protect \eqref {eq:gauss_law_map_ladder} we also know how to treat static matter. Since it can be viewed as a violation of Gauss law, we just have to change the phases of $c^{\uparrow }_i$ and $c^{\downarrow }_i$.}\BibitemShut {Stop}%
\bibitem [{\citenamefont {Fendley}(2012)}]{fendley2012parafermions}%
  \BibitemOpen
  \bibfield  {author} {\bibinfo {author} {\bibfnamefont {P.}~\bibnamefont {Fendley}},\ }\href {\doibase 10.1088/1742-5468/2012/11/p11020} {\bibfield  {journal} {\bibinfo  {journal} {Journal of Statistical Mechanics: Theory and Experiment}\ }\textbf {\bibinfo {volume} {2012}},\ \bibinfo {pages} {P11020} (\bibinfo {year} {2012})}\BibitemShut {NoStop}%
\bibitem [{\citenamefont {Whitsitt}\ \emph {et~al.}(2018)\citenamefont {Whitsitt}, \citenamefont {Samajdar},\ and\ \citenamefont {Sachdev}}]{whitsitt2018clock}%
  \BibitemOpen
  \bibfield  {author} {\bibinfo {author} {\bibfnamefont {S.}~\bibnamefont {Whitsitt}}, \bibinfo {author} {\bibfnamefont {R.}~\bibnamefont {Samajdar}}, \ and\ \bibinfo {author} {\bibfnamefont {S.}~\bibnamefont {Sachdev}},\ }\href {\doibase 10.1103/PhysRevB.98.205118} {\bibfield  {journal} {\bibinfo  {journal} {Phys. Rev. B}\ }\textbf {\bibinfo {volume} {98}},\ \bibinfo {pages} {205118} (\bibinfo {year} {2018})}\BibitemShut {NoStop}%
\bibitem [{Note2()}]{Note2}%
  \BibitemOpen
  \bibinfo {note} {For the sector $N-n$ we have that the overall factor $\cos (\pi (N-n)/N)$ is just $-\cos (\pi n/N)$. The minus sign can then be again absorbed into the $Z$'s operators. This overall operation is equivalent to the mapping $Z \DOTSB \mapstochar \rightarrow \omega ^{-n/2} Z$ for the sector $N-n$.}\BibitemShut {Stop}%
\bibitem [{\citenamefont {Schollwöck}(2011)}]{schollwock2011dmrg}%
  \BibitemOpen
  \bibfield  {author} {\bibinfo {author} {\bibfnamefont {U.}~\bibnamefont {Schollwöck}},\ }\href {\doibase https://doi.org/10.1016/j.aop.2010.09.012} {\bibfield  {journal} {\bibinfo  {journal} {Annals of Physics}\ }\textbf {\bibinfo {volume} {326}},\ \bibinfo {pages} {96} (\bibinfo {year} {2011})},\ \bibinfo {note} {january 2011 Special Issue}\BibitemShut {NoStop}%
\bibitem [{\citenamefont {Kaplan}\ and\ \citenamefont {Stryker}(2020)}]{kaplan2020gauss}%
  \BibitemOpen
  \bibfield  {author} {\bibinfo {author} {\bibfnamefont {D.~B.}\ \bibnamefont {Kaplan}}\ and\ \bibinfo {author} {\bibfnamefont {J.~R.}\ \bibnamefont {Stryker}},\ }\href {\doibase https://doi.org/10.1103/PhysRevD.102.094515} {\bibfield  {journal} {\bibinfo  {journal} {Physical Review D}\ }\textbf {\bibinfo {volume} {102}},\ \bibinfo {pages} {094515} (\bibinfo {year} {2020})}\BibitemShut {NoStop}%
\bibitem [{\citenamefont {Baxter}(2016)}]{baxter1982exactlysm}%
  \BibitemOpen
  \bibfield  {author} {\bibinfo {author} {\bibfnamefont {R.~J.}\ \bibnamefont {Baxter}},\ }\href@noop {} {\emph {\bibinfo {title} {Exactly solved models in statistical mechanics}}}\ (\bibinfo  {publisher} {Elsevier},\ \bibinfo {year} {2016})\BibitemShut {NoStop}%
\bibitem [{\citenamefont {Ba\~nuls}\ \emph {et~al.}(2011)\citenamefont {Ba\~nuls}, \citenamefont {Cirac},\ and\ \citenamefont {Hastings}}]{banuls2011thermalization}%
  \BibitemOpen
  \bibfield  {author} {\bibinfo {author} {\bibfnamefont {M.~C.}\ \bibnamefont {Ba\~nuls}}, \bibinfo {author} {\bibfnamefont {J.~I.}\ \bibnamefont {Cirac}}, \ and\ \bibinfo {author} {\bibfnamefont {M.~B.}\ \bibnamefont {Hastings}},\ }\href {\doibase 10.1103/PhysRevLett.106.050405} {\bibfield  {journal} {\bibinfo  {journal} {Phys. Rev. Lett.}\ }\textbf {\bibinfo {volume} {106}},\ \bibinfo {pages} {050405} (\bibinfo {year} {2011})}\BibitemShut {NoStop}%
\bibitem [{\citenamefont {Kormos}\ \emph {et~al.}(2017)\citenamefont {Kormos}, \citenamefont {Collura}, \citenamefont {Tak{\'a}cs},\ and\ \citenamefont {Calabrese}}]{kormos2017confinement}%
  \BibitemOpen
  \bibfield  {author} {\bibinfo {author} {\bibfnamefont {M.}~\bibnamefont {Kormos}}, \bibinfo {author} {\bibfnamefont {M.}~\bibnamefont {Collura}}, \bibinfo {author} {\bibfnamefont {G.}~\bibnamefont {Tak{\'a}cs}}, \ and\ \bibinfo {author} {\bibfnamefont {P.}~\bibnamefont {Calabrese}},\ }\href {\doibase 10.1038/nphys3934} {\bibfield  {journal} {\bibinfo  {journal} {Nature Physics}\ }\textbf {\bibinfo {volume} {13}},\ \bibinfo {pages} {246} (\bibinfo {year} {2017})}\BibitemShut {NoStop}%
\bibitem [{\citenamefont {Pomponio}\ \emph {et~al.}(2022)\citenamefont {Pomponio}, \citenamefont {Werner}, \citenamefont {Zarand},\ and\ \citenamefont {Takacs}}]{pomponio2022bloch}%
  \BibitemOpen
  \bibfield  {author} {\bibinfo {author} {\bibfnamefont {O.}~\bibnamefont {Pomponio}}, \bibinfo {author} {\bibfnamefont {M.~A.}\ \bibnamefont {Werner}}, \bibinfo {author} {\bibfnamefont {G.}~\bibnamefont {Zarand}}, \ and\ \bibinfo {author} {\bibfnamefont {G.}~\bibnamefont {Takacs}},\ }\href {\doibase 10.21468/SciPostPhys.12.2.061} {\bibfield  {journal} {\bibinfo  {journal} {SciPost Phys.}\ }\textbf {\bibinfo {volume} {12}},\ \bibinfo {pages} {61} (\bibinfo {year} {2022})}\BibitemShut {NoStop}%
\bibitem [{\citenamefont {Pfeuty}(1970)}]{pfeuty1970ising}%
  \BibitemOpen
  \bibfield  {author} {\bibinfo {author} {\bibfnamefont {P.}~\bibnamefont {Pfeuty}},\ }\href {\doibase https://doi.org/10.1016/0003-4916(70)90270-8} {\bibfield  {journal} {\bibinfo  {journal} {Annals of Physics}\ }\textbf {\bibinfo {volume} {57}},\ \bibinfo {pages} {79} (\bibinfo {year} {1970})}\BibitemShut {NoStop}%
\bibitem [{\citenamefont {Weinberg}\ and\ \citenamefont {Bukov}(2017)}]{weinberg2017quspin}%
  \BibitemOpen
  \bibfield  {author} {\bibinfo {author} {\bibfnamefont {P.}~\bibnamefont {Weinberg}}\ and\ \bibinfo {author} {\bibfnamefont {M.}~\bibnamefont {Bukov}},\ }\href {\doibase 10.21468/SciPostPhys.2.1.003} {\bibfield  {journal} {\bibinfo  {journal} {SciPost Phys.}\ }\textbf {\bibinfo {volume} {2}},\ \bibinfo {pages} {003} (\bibinfo {year} {2017})}\BibitemShut {NoStop}%
\bibitem [{\citenamefont {Weinberg}\ and\ \citenamefont {Bukov}(2019)}]{weinberg2019quspin}%
  \BibitemOpen
  \bibfield  {author} {\bibinfo {author} {\bibfnamefont {P.}~\bibnamefont {Weinberg}}\ and\ \bibinfo {author} {\bibfnamefont {M.}~\bibnamefont {Bukov}},\ }\href {\doibase 10.21468/SciPostPhys.7.2.020} {\bibfield  {journal} {\bibinfo  {journal} {SciPost Phys.}\ }\textbf {\bibinfo {volume} {7}},\ \bibinfo {pages} {20} (\bibinfo {year} {2019})}\BibitemShut {NoStop}%
\bibitem [{\citenamefont {Fishman}\ \emph {et~al.}(2022{\natexlab{a}})\citenamefont {Fishman}, \citenamefont {White},\ and\ \citenamefont {Stoudenmire}}]{itensor}%
  \BibitemOpen
  \bibfield  {author} {\bibinfo {author} {\bibfnamefont {M.}~\bibnamefont {Fishman}}, \bibinfo {author} {\bibfnamefont {S.~R.}\ \bibnamefont {White}}, \ and\ \bibinfo {author} {\bibfnamefont {E.~M.}\ \bibnamefont {Stoudenmire}},\ }\href {\doibase 10.21468/SciPostPhysCodeb.4} {\bibfield  {journal} {\bibinfo  {journal} {SciPost Phys. Codebases}\ ,\ \bibinfo {pages} {4}} (\bibinfo {year} {2022}{\natexlab{a}})}\BibitemShut {NoStop}%
\bibitem [{\citenamefont {Fishman}\ \emph {et~al.}(2022{\natexlab{b}})\citenamefont {Fishman}, \citenamefont {White},\ and\ \citenamefont {Stoudenmire}}]{itensor-r0.3}%
  \BibitemOpen
  \bibfield  {author} {\bibinfo {author} {\bibfnamefont {M.}~\bibnamefont {Fishman}}, \bibinfo {author} {\bibfnamefont {S.~R.}\ \bibnamefont {White}}, \ and\ \bibinfo {author} {\bibfnamefont {E.~M.}\ \bibnamefont {Stoudenmire}},\ }\href {\doibase 10.21468/SciPostPhysCodeb.4-r0.3} {\bibfield  {journal} {\bibinfo  {journal} {SciPost Phys. Codebases}\ ,\ \bibinfo {pages} {4}} (\bibinfo {year} {2022}{\natexlab{b}})}\BibitemShut {NoStop}%
\bibitem [{\citenamefont {Schwinger}\ and\ \citenamefont {Englert}(2001)}]{schwinger2001symbolism}%
  \BibitemOpen
  \bibfield  {author} {\bibinfo {author} {\bibfnamefont {J.}~\bibnamefont {Schwinger}}\ and\ \bibinfo {author} {\bibfnamefont {B.-G.}\ \bibnamefont {Englert}},\ }\href@noop {} {\emph {\bibinfo {title} {Symbolism of Atomic Measurements}}}\ (\bibinfo  {publisher} {Springer, Berlin},\ \bibinfo {year} {2001})\BibitemShut {NoStop}%
\end{thebibliography}%

\end{document}